\documentclass[a4paper,10pt,prc,twocolumn,showpacs,nofootinbib]{revtex4}
\usepackage{graphicx}
\usepackage{amsfonts}
\usepackage{amsmath}
\usepackage{amsbsy}
\usepackage{hyperref}
\renewcommand{\vec}[1]{\mathbf{#1}}
\def\be{\begin{equation}}
\def\ee{\end{equation}}
\def\bea{\begin{eqnarray}}
\def\eea{\end{eqnarray}}

\begin{document}

\title{Numerical tests of causal relativistic
dissipative fluid dynamics}

\bigskip

\author{E.\ Moln\'ar$^1$, H.\ Niemi$^1$, and
D.H.\ Rischke$^{1,2}$}
\affiliation{$^1$Frankfurt Institute for Advanced Studies,
Ruth-Moufang-Str.\ 1, D-60438 Frankfurt am Main, Germany\\
$^2$ Institut f\"ur Theoretische Physik,
Johann Wolfgang Goethe-Universit\"at,
Max-von-Laue-Str.\ 1, D-60438 Frankfurt am Main, Germany
}

\bigskip



\date{\today}

\begin{abstract}
We present numerical methods to solve the Israel-Stewart (IS)
equations of causal relativistic dissipative fluid dynamics
with bulk and shear viscosities. We then test these methods
studying the Riemann problem in (1+1)-- and (2+1)-dimensional
geometry. The numerical schemes investigated here are
applicable to realistic (3+1)--dimensional modeling of a
relativistic dissipative fluid.
\end{abstract}

\maketitle

\section{Introduction}

The interest in modeling the evolution of matter created
in relativistic heavy-ion collisions with fluid dynamics has
never ceased since the pioneering works by Landau
\cite{Belenkij:1956cd}. Recent remarkable discoveries at the
Relativistic Heavy Ion Collider (RHIC) at Brookhaven National
Laboratory provide evidence for an almost ``perfect''
fluid-like behavior of the QCD matter created \cite{RHIC_PF}.

In a perfect, or ideal, fluid transport coefficients
like bulk and shear viscosity and heat conductivity
vanish. This is an idealized situation; in a real fluid
one can show that there are lower bounds for these transport
coefficients, for instance using the uncertainty principle
\cite{Danielewicz:1984ww} or applying the
AdS/CFT conjecture \cite{Kovtun:2004de}.
In order to decide how close to a perfect fluid
the matter created at RHIC is,
one has perform theoretical calculations in the framework
of causal relativistic dissipative fluid dynamics.
In this way, one may also be able to extract the numerical
values for the bulk and shear viscosity coefficients
from experimental measurements.

Currently the most widely accepted and studied theory
of relativistic dissipative fluid dynamics is due to
Israel and Stewart \cite{Stewart:1972hg,Israel:1976tn,
Stewart:1977,Israel:1979wp}.
This is the relativistic version of the pioneering work
by M\"uller \cite{Muller,Muller_2}. Although these theories
have been developed in the 1970's, efforts to study
and apply them to relativistic heavy-ion collisions
have only started very recently
\cite{Muronga:2001zk,Muronga:2003ta}.
This has been followed by an impressive number of
studies in (1+1)--dimensional
\cite{BRW,Dumitru:2007qr,Huovinen:2008te,Baier:2006gy}
and (2+1)--dimensional geometries
\cite{Muronga:2004sf,Chaudhuri:2007zm,
Romatschke:2007mq,Chaudhuri:2008sj,Luzum:2008cw,
Song:2007fn,Song:2007ux,Song:2008si}.

In 3+1 dimensions, given arbitrary initial conditions
and a general equation of state, the only way to solve the
equations of relativistic fluid dynamics is by means of
numerical methods. Any numerical method requires an
algorithm that has to be tested in order to
assess its validity for solving the underlying equations.
Testing algorithms to solve relativistic dissipative
fluid dynamics is made difficult by the fact that there
is only a rather limited number of test cases with
analytical solutions. Reference \cite{Baier:2006gy}
investigated sound propagation for the linearized IS equations.
The algorithm of Ref.\ \cite{Song:2007ux} was checked,
for certain expansion scenarios, as to whether it
correctly approaches the Navier-Stokes
and ideal-fluid limits. So far, however, numerical algorithms
to solve the IS equations have not been tested in situations
where shock discontinuities occur in the ideal-fluid limit.
The present paper, in which we perform
an extensive study of the relativistic Riemann problem
in 1+1 and 2+1 dimensions, aims to fill this gap.

In Sec.\ \ref{II} we provide a
short review of IS theory of dissipative fluid dynamics.
In Sec.\ \ref{III} we formulate it in a form suitable for
numerical implementation. This is followed in Sec.\ \ref{IV}
by an introductory presentation to the numerical methods
which we use to solve the IS equations. In Sec.\ \ref{V}
we report results of solving the Riemann problem
in 1+1 and 2+1 dimensions.
Section \ref{VI} concludes this work with a summary of
our results and an outlook.

\section{Dissipative fluid dynamics} \label{II}
\subsection{Units and definitions}

Throughout this work natural units, $\hbar = c = k_B = 1$,
are used. Components of contravariant vectors and tensors
in 4-dimensional space-time are denoted by upper indices, i.e.,
$A^{\mu}$ and $A^{\mu\nu}$. Greek indices take values
from $0$ to $3$ and Roman indices from $1$ to $3$.
Covariant components, denoted by lower indices, are obtained by
$A_{\nu} \equiv  g_{\mu \nu} A^{\mu}$,
where $g_{\mu\nu}$ is the metric tensor,
for which we use the $(+, -, -, -)$ convention.
If not stated otherwise the Einstein summation convention
is used for both Greek and Roman indices.

For an arbitrary contravariant four-vector
$A^{\mu}$ the covariant derivative is defined as
\bea
A^{\mu}_{;\alpha} &\equiv& \partial_{\alpha}A^{\mu}
+ \Gamma^{\mu}_{\alpha \beta} A^{\beta} \, ,
\eea
where
$\Gamma^{\mu}_{\alpha \beta} \equiv
\frac{1}{2} \, g^{\mu \nu} \left(\partial_{\beta} g_{\alpha \nu}
+ \partial_{\alpha} g_{\nu \beta}
- \partial_{\nu} g_{\alpha \beta} \right)$
denotes the Christoffel symbol of the second kind and
$\partial_{\alpha} = \partial/\partial x^\alpha$ denotes
the four-derivative. Similarly, the covariant
derivative of covariant vectors is given by
$A_{\mu;\alpha} \equiv \partial_{\alpha}A_{\mu} -
\Gamma^{\beta}_{\mu \alpha} A_{\beta}$.
For scalars the covariant derivative reduces to the
ordinary four-derivative.
The covariant derivative of second-rank contravariant tensors is
\bea
A^{\mu \nu}_{;\alpha} &\equiv& \partial_{\alpha} A^{\mu \nu}
+ \Gamma^{\mu}_{\alpha \beta} A^{\beta \nu}
+ \Gamma^{\nu}_{\alpha \beta} A^{\mu \beta}\,.
\eea
Vectors and tensors can be decomposed into parts parallel
and orthogonal to the four-velocity of matter $u^\mu$,
where $u^\mu u_\mu = 1$.
Using the transverse projection operator
$\Delta^{\mu\nu} \equiv g^{\mu \nu} - u^{\mu} u^{\nu}$
where $\Delta^{\mu \nu} u_{\nu} = 0$,
an arbitrary four-vector can be written as
$A^\mu = u^\mu u_\alpha A^\alpha + \Delta^{\mu\alpha}A_\alpha$.
The covariant derivative of an arbitrary tensor can be
decomposed as
\be
A^{\mu_1 \dots \mu_n}_{;\alpha} \equiv
u_\alpha D A^{\mu_1 \dots \mu_n}
+ \nabla_\alpha A^{\mu_1 \dots \mu_n}\,,
\ee
where the convective time derivative $D$ and the spatial
gradient operator $\nabla_\alpha$ are given by
\bea
D A^{\mu_1 \dots \mu_n} &\equiv& u^{\beta}
A^{\mu_1 \dots \mu_n}_{;\beta} \, , \label{timeD}\\
\nabla_\alpha A^{\mu_1 \dots \mu_n} &\equiv&
\Delta_{\alpha}^{\beta} A^{\mu_1 \dots \mu_n}_{;\beta}\,.
\label{grad}
\eea
It is convenient to define the traceless and
symmetric projection of a tensor field,
which is orthogonal to $u^\mu$.
This is denoted by angular brackets $\langle \rangle$,
\bea
A^{\langle \mu \nu \rangle} \equiv
\frac{1}{2} \Delta^{\mu\alpha}\Delta^{\nu\beta}
(A_{\alpha\beta} + A_{\beta\alpha})
- \frac{1}{3} \Delta^{\mu\nu} \Delta_{\alpha\beta}
A^{\alpha\beta} \, .
\eea
The covariant derivative of the four-velocity can be generally
decomposed as
\begin{equation}
 u_{\nu;\mu} = u_\mu Du_\nu + \sigma_{\mu\nu}
+ \frac{1}{3}\Delta_{\mu \nu}\theta - \omega_{\mu\nu},
\end{equation}
where the expansion rate $\theta$,
the shear tensor $\sigma_{\mu\nu}$, and the
vorticity tensor $\omega_{\mu\nu}$ are defined as
\bea
\label{theta}
\theta &\equiv& \nabla_\mu u^\mu =
\partial_{\mu} u^{\mu} + \Gamma^{\mu}_{\alpha \mu} u^{\alpha}
\, , \\  \nonumber
\sigma^{\mu \nu} &\equiv& \nabla^{<\mu} u^{\nu >}
=\frac{1}{2} \Delta^{\mu\alpha}\Delta^{\nu\beta}
(u_{\alpha;\beta} + u_{\beta;\alpha})
- \frac{\theta}{3} \Delta^{\mu\nu} \\ \nonumber
& = & \frac{1}{2}\left(\partial^\mu u^\nu - u^{\mu} u^\alpha
\partial_\alpha u^{\nu}
+ \partial^\nu u^\mu - u^{\nu} u^\alpha
\partial_\alpha  u^{\mu}\right) \\
&+& \frac{1}{2} \left(\Delta^{\mu \alpha} u^{\beta}
\Gamma^{\nu}_{\alpha \beta} + \Delta^{\nu \alpha} u^{\beta}
\Gamma^{\mu}_{\alpha \beta} \right)
 - \frac{\theta}{3}\Delta^{\mu \nu},  \label{shear} \\
\omega^{\mu}_{\hspace*{0.1cm}\nu} &\equiv&
\frac{1}{2}\Delta^{\mu \alpha}
\Delta^{\beta}_{\hspace*{0.1cm}\nu}
\left(u_{\alpha;\beta} - u_{\beta;\alpha} \right)
\\ \nonumber
&=& \frac{1}{2} \left( \frac{}{}
\partial_{\nu}u^{\mu} - \partial^{\mu} u_{\nu} + u^{\mu} u^\alpha
\partial_\alpha u_{\nu} - u_{\nu} u^\alpha
\partial_\alpha u^{\mu} \right)\,.\label{vorticity}
\eea
where $\sigma^{\mu\nu} u_{\nu} = 0$
and $\omega^{\mu\nu} u_{\nu} = 0$.

\subsection{The equations of causal relativistic dissipative fluid dynamics}

The basic quantities characterizing dissipative fluids
are the net charge current $N^\mu$ and the energy-momentum
tensor $T^{\mu\nu}$. Following Refs.\
\cite{Eckart:1940te,Landau_book,DeGroot,Csernai_book}
these can be decomposed with respect to
the fluid four-velocity $u^\mu$ as
\bea\label{N_mu}
N^{\mu} &\equiv& N^{\mu}_{eq} + \delta N^{\mu} =
n u^{\mu} + V^{\mu}\, , \\ \nonumber
T^{\mu \nu}
&\equiv& T^{\mu \nu}_{eq} + \delta T^{\mu \nu}
= e u^{\mu} u^{\nu} - (p + \Pi)\Delta^{\mu \nu}\\  \label{T_munu}
&+& W^{\mu}u^{\nu} + W^{\nu}u^{\mu} + \pi^{\mu \nu}\,,
\eea
where $n \equiv N^{\mu}u_{\mu}$ is the net charge density
and $e \equiv u_{\mu} T^{\mu \nu} u_{\nu}$ is the energy density
in the local rest frame (LRF), i.e., where $u^{\mu} = (1,0,0,0)$.
The charge diffusion current is given by
$\delta N^{\mu} \equiv V^{\mu} = N_{\nu} \Delta^{\mu \nu}$.
The energy-momentum flow orthogonal to $u^\mu$ is given by
$W^{\mu} \equiv \Delta^{\mu \alpha} T_{\alpha \beta} u^{\beta}$.
This quantity can be decomposed as
$W^{\mu} \equiv q^{\mu} + (e + p)V^{\mu}/n$, where
$q^\mu$ is the heat flow.
The local isotropic pressure is denoted by
$p + \Pi \equiv -\frac{1}{3}\Delta_{\mu \nu} T^{\mu \nu}$,
where $p$ is the equilibrium pressure and $\Pi$ is
the bulk viscous pressure measuring the deviation from the
local equilibrium pressure. The shear stress tensor is
defined as $\pi^{\mu \nu} \equiv T^{\langle \mu\nu \rangle}$.
This representation is completely general, valid in any
coordinate system, and independent of
the definition of the flow velocity.

Usually, there are two typical choices used to define the 
flow velocity: either tied to the net charge flow when $V^\mu = 0$
(Eckart frame) or tied to the energy flow
when $W^\mu = 0$ (Landau frame).
We will use the latter definition in this work.

Without conserved charges only Landau's definition of the
flow velocity is appropriate.
In this case the heat flow is $q^\mu = - (e+p)V^\mu/n$.
For net charge-free matter, $q^\mu$ is not well defined,
but also irrelevant for the discussion, so we set it
to zero, $q^\mu = V^\mu = 0$.

When all dissipative quantities are zero,
$V^\mu = W^\mu = \Pi = \pi^{\mu \nu} = 0$, the decompositions
\eqref{N_mu} and \eqref{T_munu} reduce to
perfect fluid form,
$N^{\mu} = N_{eq}^{\mu} \equiv n u^{\mu}$
and $T^{\mu \nu} = T_{eq}^{\mu \nu} \equiv
e u^{\mu} u^{\nu} - p(e,n)\Delta^{\mu \nu}$.
The LRF energy and charge densities are always
fixed to their equilibrium values by the Landau 
matching conditions, i.e., $n=n_{eq}$, and $e=e_{eq}$.
Then, the equilibrium pressure is given by the
equation of state (EOS) $p = p(e, n)\equiv
-\frac{1}{3}\Delta_{\mu \nu} T^{\mu \nu}_{eq}$.

The equations of relativistic dissipative fluid dynamics
follow from the covariant differentiation of the
conserved charge four-current and the energy-momentum tensor,
\bea
N^{\mu}_{;\mu} &\equiv& \frac{1}{\sqrt{g}}\,
\partial_\mu \left(\sqrt{g} \, N^{\mu} \right) = 0\,
, \label{cons1} \\
T^{\mu \nu}_{;\mu} &\equiv& \frac{1}{\sqrt{g}}\,
\partial_\mu \left(\sqrt{g}  \, T^{\mu \nu} \right)
+ \Gamma^{\nu}_{\mu \beta} T^{\mu \beta}  = 0\,
, \label{cons2}
\eea
where $g \equiv -\textrm{det}(g_{\mu \nu})$ is the
negative determinant of the metric tensor.

The non-equilibrium entropy current can be written as
\bea
S^{\mu} &\equiv& S^{\mu}_{eq} + \delta S^{\mu}
= s u^{\mu} + \Phi^{\mu}\, , \label{S_mu}
\eea
where the entropy flux relative to
$u^\mu$ is $\Phi^{\mu}
= S_{\nu} \Delta^{\mu \nu}$.
The LRF entropy density is
$s = S^{\mu} u_{\mu}$, where in general $s \leq s_{eq}(e,n)$.

Following Refs.\
\cite{Stewart:1972hg,Israel:1976tn,Stewart:1977,Israel:1979wp},
the phenomenological extension of the entropy
four-current by Israel and Stewart can be written
without heat conductivity as
\bea\label{IS_entropy}
S^{\mu} \equiv s u^{\mu} = s_{eq} u^{\mu}
- \left(\beta_0 \Pi^2 +  \beta_2 \pi^{\alpha \beta}
\pi_{\alpha \beta}  \right)\frac{u^{\mu}}{2 T} \, ,
\eea
where the coefficients $\beta_0,\beta_2$  are functions
of $e$ and $n$. Their exact value can be
determined explicitly e.g.\ from kinetic theory.

The requirement of non-decreasing entropy leads
to relaxation equations for the bulk pressure and shear stress tensor.
Here we also include the vorticity terms which follow
from the kinetic-theory derivation, but we neglect the
coupling between bulk and shear viscosity. Then,
the IS equations \cite{Israel:1979wp,Huovinen:2008te} read
\bea\label{relax_bulk}
D\Pi &=&  \frac{1}{\tau_\Pi}\left(\Pi_{NS} - \Pi \right)
- I_0 \, ,\\
\label{relax_shear}
D \pi^{\mu\nu} &=& \frac{1}{\tau_\pi}
\left(\pi^{\mu\nu}_{NS} - \pi^{\mu\nu} \right)
- I^{\mu \nu}_1  - I^{\mu \nu}_2 - I^{\mu \nu}_3 \, ,
\eea
where $\tau_{\Pi} = \zeta \beta_0$ denotes the
relaxation time of the bulk viscous pressure
and $\tau_{\pi} =  2\eta \beta_2$ is the relaxation time
of the shear stress tensor. The relativistic
Navier-Stokes values are given by
\cite{Eckart:1940te,Landau_book}
\bea
\Pi_{NS} &\equiv& -\zeta \theta \, ,  \label{bulk_NS}\\
\pi^{\mu \nu}_{NS} &\equiv& 2 \eta \sigma^{\mu \nu}
\label{shear_NS} \, ,
\eea
where $\zeta \geq 0$ is the bulk viscosity coefficient and
$\eta \geq 0$ is the shear viscosity coefficient.
In Eqs.\ (\ref{relax_bulk}), (\ref{relax_shear}),
we introduced the abbreviations
\bea\label{I_0}
I_0 & \equiv & \frac{1}{2}\Pi \left(\nabla_\lambda u^\lambda
+ D \ln \frac{\beta_0}{T}\right)\, ,\\
\label{I_1}
I^{\mu \nu}_1 &\equiv& (\pi^{\lambda\mu}u^\nu
+\pi^{\lambda\nu}u^\mu)D u_\lambda \, , \\
I^{\mu \nu}_2 &\equiv& \frac{1}{2}\pi^{\mu\nu}
\left(\nabla_\lambda u^\lambda
+ D \ln \frac{\beta_2}{T}\right) \, ,
\label{I_2}\\
I^{\mu\nu}_3 & \equiv&  2 \pi_\lambda^{\langle\mu}
\omega^{\nu\rangle\lambda}
= \pi^{\mu\lambda} \omega^{\nu}_{\hspace*{0.1cm} \lambda}
+ \pi^{\nu \lambda} \omega^{\mu}_{\hspace*{0.1cm} \lambda} \, ,
\eea
where we used $\pi^{\mu\nu} \omega_{\mu \nu} = 0$.

For the sake of simplicity, in our numerical studies
presented in the subsequent sections we assume
a gas of massless Boltzmann particles without conserved charges.
In this case, the equation of state is simply $e = 3p$
and $e = \frac{3g}{\pi^2} T^4$,
where $g$ is the number of degrees of freedom.
The equilibrium entropy density is given by
$s_{eq} = \frac{4g}{\pi^2} T^3$.
In this case, we can further simplify
$I^{\mu \nu}_2$ noting that the exact value
of the thermodynamic integral for massless Boltzmann gas is,
$\beta_2 = 3/(4p)$ \cite{Israel:1979wp,Muronga:2006zx}.
Therefore, it follows that $D \beta_2/\beta_2 = - D e/e$.
Thus, $D \ln \left(\beta_2/T\right) = - De/e - DT/T$,
where the temperature can be calculated from the EOS.
The convective time derivative of the LRF energy density
is given by energy conservation,
$De = -(e + P) u^{\mu}_{;\mu} - \pi^{\mu \nu} u_{\mu;\nu}$,
where the effective pressure $P$ is defined as
$P(e,n,\Pi) = p(e,n) + \Pi$.

\subsection{General coordinate representation}

Here we give the relations between $T^{\mu\nu}$ and
$N^\mu$ in the calculational, or laboratory, frame
and the LRF densities $e$, $n$, and the flow velocity $v^i$.
The natural frame of reference is the laboratory
frame. However, during the time evolution
of the system, we have to extract the local velocity
and the LRF densities from the laboratory frame quantities.
These are needed because the EOS is given as a
function of LRF densities, $p=p(e,n)$.

We can write the four-vector and tensor quantities
given in Eqs.\ (\ref{N_mu}), (\ref{T_munu})
by specifying the four-velocity of the matter,
$u^{\mu} = \gamma(1,v_i) = \gamma(1,v_x,v_y,v_z)$, where
$\gamma=(1 - v^2)^{-1/2}$ and $v \equiv |\vec{v}|
= (v_x^2 + v_y^2 + v_z^2)^{1/2}$.
The laboratory frame quantities take the form
\bea\label{n0}
N^{0} &\equiv& n\gamma \, ,\\  \label{ni}
N^{i} &\equiv& n\gamma v_i = v_i N^{0}\, ,\\  \label{t00}
T^{00} &\equiv& (e + P)\gamma^2 - g^{00}P + \pi^{00} \, , \\
\nonumber
T^{0i} &\equiv& (e + P )\gamma^2 v_i - g^{0i} P  + \pi^{0i} \, ,
\\ \label{t0i}
&=& v_i T^{00} + P (g^{00} v_i - g^{0i}) - v_i \pi^{00}
+ \pi^{0i} \, , \\ \nonumber
T^{ij} &\equiv& (e + P)\gamma^2 v_iv_j  - P g^{ij} +
\pi^{ij} \, ,\\ \label{tij}
&=& v_i T^{0j} + P (g^{0j}v_i - g^{ij}) -
v_i \pi^{0j} + \pi^{ij} \, .
\eea
$N^{0}$ is the local charge density,
$N^{i}$ is the local charge current in the direction $i$,
i.e., the direction of the flow $u^i$.
The total energy density of the fluid is $T^{00}$
which in the LRF reduces to the (equilibrium) energy density $e$.
By definition, $T^{0i}$ denotes the energy flow in the
direction of $u^i$, while $T^{i0}$ is the momentum density
flux in the $i$th direction\footnote{In standard units
the flow of the energy density is $c T^{0i}$,
while the flow of momentum density is $c^{-1} T^{i0}$.}.
The remaining spatial part, $T^{ij}$, denotes the
$i$th component of the momentum flowing in direction $j$.

The LRF charge density and energy density are obtained
from Eq.\ (\ref{n0}) and Eqs.\ (\ref{t00}), (\ref{t0i}),
respectively,
\bea
n &=& N^{0} (1 - v^2)^{1/2} \, , \\ \label{e_density}
e &=& T^{00} - \pi^{00} - v_i (T^{0i} - \pi^{0i}) \, ,
\eea
while Eq.\ (\ref{t0i}) together with the above expressions
leads to the expression for the velocity components,
\bea\label{velocity}
v_i = \frac{T^{0i} - \pi^{0i} +
P g^{0i}}{T^{00} - \pi^{00} + P g^{00}} \, .
\eea
In most cases of interest $g^{00} = 1$, and
the metric of the space-time is diagonal.
Therefore we can introduce a simplified notation which mimics the
perfect fluid relations~\cite{Heinz:2005bw}, $R = n\gamma$,
$E \equiv T^{00} - \pi^{00}$, $M_i \equiv T^{0i} - \pi^{0i}$,
where $M \equiv |\vec{M}| = (M_x^2 + M_y^2 + M_z^2)^{1/2}$.
Thus, $\vec{M}$ is parallel to the velocity $\vec{v}$,
similarly as in the perfect fluid case.
These quantities have to obey the physical constraint 
$M\leq E$, in order to obtain meaningful solutions.
Therefore, we can express the LRF charge density,
energy density, the absolute magnitude
of the velocity, and the velocity components as
\bea \label{n_lr}
n &=& R(1 - \vec{v} \cdot \vec{v})^{1/2}\, ,\\
e &=& E - \vec{v} \cdot \vec{M} \, , \label{e_lr}\\
v &=& M/\left[E + P\right]\, , \label{velo_magnitude} \\
v_i &=& v M_i/M \, . \label{velo_components}
\eea
Substituting Eqs.\ (\ref{n_lr}), (\ref{e_lr})
into Eq.\ (\ref{velo_magnitude}) we
obtain the equation for the magnitude of the velocity, $v$.
This can be solved by using a one-dimensional root search.
Thereafter, use of Eq.\ (\ref{velo_components}) yields
the individual velocity components and $\gamma$.
Note that, in case of perfect fluids, this simplified treatment is
practicable, however, in case of dissipative fluids, this may not
always be possible.
This is due to the fact that the vectors $T^{0 \mu}$ and $\pi^{0 \mu}$ 
are not parallel to each other.
Hence choosing other shear stress tensor components as independent
variables, or in cases which take into account the heat flow,
it is required to carry out a multidimensional root search to find the velocity
\cite{Heinz:2005bw,Muronga:2006zw,Chaudhuri:2007zm,Chaudhuri:2008sj}.

For dissipative fluids the number of unknown variables
increases by the introduction of the shear stress tensor
and the bulk viscosity. The shear stress tensor is constrained
by the orthogonality condition
$\pi^{\mu \nu} u_{\nu} = 0$, leading to the following relations,
\bea\label{pi0i}
\pi^{i0}u_0 &\equiv& - \pi^{ij} u_j \, ,\\
\pi^{00}u_0 &\equiv& -\pi^{0i} u_i = \pi^{ij} u_j u_i/u_0 \, .
\label{pi001}
\eea
One more independent relation follows from the trace of
the shear stress tensor, $\pi^{\mu}_{\mu} = 0$,
\bea\label{pi002}
\pi^{00} \equiv -\pi^{ii}g_{ii} \, .
\eea
In general, using Eq.\ (\ref{pi0i}) we can reduce the
number of unknowns by three, and by two using
Eqs.\ (\ref{pi001}) and (\ref{pi002}).
Thus we are left with five independent components of the
shear stress tensor. However, for testing the numerical
solutions it is preferable to calculate all shear stress
tensor components directly using the relaxation equations,
instead of using the orthogonality relations.
We will return to this matter later and provide examples.

\section{Test problems} \label{III}

In this section we shall write the IS equations
in various (1+1)-- and (2+1)--dimensional geometries with
Cartesian or curvilinear coordinates.
For the sake of completeness, the (2+1)--dimensional
boost-invariant and the (3+1)--dimensional IS equations
in Cartesian as well as in  $(\tau,x,y,\eta)$
coordinates are given in the Appendices.
Here $\tau$ is the longitudinal proper time and
$\eta$ is the space-time rapidity.

\subsection{(1+1)--dimensional Cartesian coordinates}\label{1d}

In Cartesian coordinates, the metric tensor is
$g^{\mu \nu} \equiv \eta^{\mu \nu} = \textrm{diag}(1,-1,-1,-1)$
and all Christoffel symbols vanish.
The negative determinant of the metric is $g = 1$.
We assume that the system evolves along the $z$ direction and
that it is homogeneous in the transverse plane,
such that the spatial derivatives in $x$ and $y$ directions
vanish identically. The flow velocity of matter is
$u^{\mu} = \gamma_z (1,0,0,v_z)$ and
$\gamma_z = (1 - v^2_z)^{-1/2}$.

The components of the energy-momentum tensor and charge
current in the laboratory frame are
\bea
N^{0} &\equiv& n\gamma_z \, , \\
N^{z} &\equiv& N^{0} v_z \, , \\
T^{00} &\equiv& (e + P)\gamma^2_z - P + \pi^{00}
= (e + P_z)\gamma^2_z - P_z \, , \qquad\\
T^{0z} &\equiv& (e + P)\gamma^2_z v_z + \pi^{0z}
= (e + P_z)\gamma^2_z v_z\, ,\\
T^{xx} &\equiv& \pi^{xx} + P = -\frac{\pi}{2} + P\, , \\
T^{yy} &\equiv& \pi^{yy} + P = -\frac{\pi}{2} + P\, , \\
T^{zz} &\equiv& (e + P)\gamma^2_z v^2_z + P + \pi^{zz}
= (e v^2_z + P_z)\gamma^2_z\, ,
\eea
where the shear pressure, $\pi$,
is defined such that $\pi^{zz} = \gamma^2_z \pi$.
The orthogonality and tracelessness properties imply
$\pi^{xx} = \pi^{yy} = -\pi/2$, and
$\pi^{00} = v^{2}_z \gamma^2_z \pi$,
see Ref.\ \cite{Muronga:2006zw}.
From the orthogonality relation we obtain
$\pi^{0z} = v_z \pi^{zz} = v_z \gamma^2_z \pi$.
The effective pressure in the $z$ direction is denoted
by $P_z \equiv P + \pi = p(e,n) + \Pi + \pi$.
The remaining four-vector and tensor components vanish,
$N^{x} = N^{y} = 0$ and $T^{0x}=T^{0y}=T^{xy}=T^{xz}=T^{yz}=0$.
This also means that the corresponding shear stress tensor
components vanish,
$\pi^{0x}=\pi^{0y}=\pi^{xy}=\pi^{xz}=\pi^{yz}=0$.

The LRF quantities and the velocity can be expressed
in terms of the laboratory quantities,
\bea
n &=& N^{0} \left(1 - v^2_z \right)^{1/2} \, , \\
\label{eq:e_1d}
e &=& T^{00} - v_z T^{0z} \, ,\\
\label{eq:v_1d}
v_z &=& \frac{T^{0z}}{T^{00} + P_z}\, .
\eea
The conservation equations follow from
Eqs.\ (\ref{cons1}), (\ref{cons2}),
\bea\label{N01d_cons}
\partial_t N^{0} + \partial_z (v_z N^{0}) &=& 0 \, , \\
\label{T001d_cons}
\partial_t T^{00} + \partial_z (v_z T^{00}) &=&
- \partial_z (v_z P_z)\, , \\
\label{T0i1d_cons}
\partial_t T^{0z} + \partial_z (v_z T^{0z}) &=&
- \partial_z P_z \, .
\eea
The relaxation equations for the bulk viscous and
shear pressure
follow from Eqs.\ (\ref{relax_bulk}), (\ref{relax_shear}):
\bea
\gamma_z \partial_t \Pi + \gamma_z v_z \partial_z \Pi
\label{relax_bulk_1d}
&=& \frac{1}{\tau_\Pi}\left(\Pi_{NS} - \Pi \right) - I_0 \, , \\
\gamma_z \partial_t \pi + \gamma_z v_z \partial_z \pi
&=& \frac{1}{\tau_\pi}\left(\pi_{NS} - \pi \right) - I_2 \, ,
 \label{relax_shear_1d}
\eea
where $I^{xx}_1 = I^{xx}_3 = 0$.
The Navier-Stokes values of the bulk viscous
and shear pressure are
\bea
\Pi_{NS} &\equiv& - \zeta \theta_z \, , \\
\pi_{NS} &\equiv& 2\eta \sigma = -\frac{4}{3} \eta \theta_z \, ,
\eea
where $\theta_z \equiv \partial_{\mu} u^{\mu} =
\partial_t \gamma_z + \partial_z (\gamma_z v_z) =
\nabla_\mu u^{\mu}$ denotes the expansion scalar and
$\sigma = -2\sigma^{xx} = \theta_z/3$ is the shear stress.
Furthermore Eqs.\ (\ref{I_0})
and (\ref{I_2}) with $I_2 = -2 I^{xx}_2$, lead to
\bea
I_0 &=& \frac{\Pi}{2}
\left(\theta_z + D \ln \frac{\beta_0}{T}\right) \, , \\
I_2 &=& \frac{\pi}{2}
\left(\theta_z +  D \ln \frac{\beta_2}{T}\right) \, .
\eea

\subsection{(1+1)-dimensional cylindrical coordinates}
\label{1d_r}

In the case of (1+1)--dimensional cylindrical coordinates,
all quantities are functions of the time $t$ and the radial
coordinate $r$ only. The flow velocity is given by
$u^{\mu} = \gamma_r (1,v_r,0,0)$,
where $\gamma_r = (1 - v^2_r)^{-1/2}$.
The gradient operator is
$\partial_{\mu} = (\partial_t, \partial_r, 0, 0)$.
The terms containing $\partial_\phi$ and $\partial_z$
vanish identically.

The metric tensor transforms as
$g_{\mu \nu} = \frac{\partial
\tilde x^{\alpha}}{\partial x^{\mu}}
\frac{\partial \tilde x^{\beta}}{\partial x^{\nu}}
\eta_{\alpha \beta}$,
where $x^{\mu} = (t,r,\phi,z)$,
$\tilde{x}^{\mu} = (t,x,y,z)$ and
$\eta_{\mu \nu}$ is the Cartesian metric.
The spatial coordinates are $r = \sqrt{x^2 + y^2}$
and $\phi=\arctan(y/x)$.
The contravariant and covariant metric tensors are
$g^{\mu \nu} = \textrm{diag}(1,-1,-1/r^2,-1)$ and
$g_{\mu \nu} = \textrm{diag}(1,-1,-r^2,-1)$, respectively.
The negative determinant is $g = r^2$.
The only non-vanishing Christoffel symbols are
$\Gamma^{\phi}_{\phi r} = \Gamma^{\phi}_{r \phi} = r^{-1}$
and $\Gamma^{r}_{\phi \phi} = -r$.

The laboratory frame quantities are
\bea
N^{0} &\equiv& n\gamma_r \, , \\
N^{r} &\equiv& N^{0} v_r \, , \\
T^{00} &\equiv& (e + P)\gamma^2_r - P + \pi^{00}
= (e + P_r)\gamma^2_r - P_r \, , \quad \\
T^{0r} &\equiv& (e + P )\gamma^2_r v_r + \pi^{0r}
= (e + P_r)\gamma^2_r v_r \, , \\
T^{rr} &\equiv& (e + P)\gamma^2_r v^2_r + P +  \pi^{rr}\, ,
\nonumber \\
&=& (e + P_r)\gamma^2_r v^2_r + P_r \, , \\
T^{\phi\phi} &\equiv& \frac{P}{r^2} + \pi^{\phi \phi}\, , \\
T^{zz} &\equiv& P + \pi^{zz} \, ,
\eea
where $P_r$ is the effective pressure in the radial direction defined below.
All remaining vector and tensor components vanish identically,
$N^{\phi} = N^{z} = 0$,
$T^{0\phi}= T^{0z}=T^{\phi r}=T^{\phi z}=T^{rz} = 0$ and
$\pi^{0\phi} = \pi^{0z} = \pi^{\phi r} = \pi^{\phi z}
= \pi^{rz}= 0$.

To reduce the number of unknowns we use the
transversality of the shear stress tensor, leading to
$\pi^{0r} = v_r \pi^{rr}$, $\pi^{00} = v^2_r \pi^{rr}$.
The tracelessness condition gives
$\pi^{00} = \pi^{rr} + r^2\pi^{\phi \phi} + \pi^{zz}$.
The simplest solution is to choose $\pi^{\phi \phi}$
and $\pi^{zz}$ as the independent components of the shear
stress tensor, since a Lorentz boost in radial
direction does not affect these components.
The remaining shear stress tensor components
can be expressed by using these components,
\bea
\pi^{rr} &=& -\gamma^2_r (r^2 \pi^{\phi \phi} + \pi^{zz})\, , \\
\pi^{0r} &=& -v_r \gamma^2_r (r^2 \pi^{\phi \phi} + \pi^{zz})\,
, \\
\pi^{00} &=& -v^2_r \gamma^2_r (r^2 \pi^{\phi \phi} + \pi^{zz})
\, .
\eea
The LRF charge density, energy density, and velocity
are given as
\bea
n &=& N^{0} \left(1 - v^2_r \right)^{1/2} \, , \\
e &=& T^{00} - v_r T^{0r} \, , \\
v_r &=& \frac{T^{0r}}{T^{00} + P_r} \, ,
\eea
where $P_r \equiv P + \frac{\pi^{rr}}{\gamma^2_r} 
= P - r^2 \pi^{\phi \phi} - \pi^{zz}$.

The charge conservation equation and the equations of
energy and momentum conservation follow from
Eqs.\ (\ref{cons1}), (\ref{cons2}),
\bea
\partial_t N^{0} + \partial_{r}\left(v_r N^{0}\right)
&=& - \frac{1}{r}\left(v_r N^{0}\right) \, , \\
\partial_t T^{00} + \partial_{r}\left(v_r T^{00}\right)
&=&  - \partial_{r}\left(v_r P_r \right) \nonumber \\
&-& \frac{1}{r}\left(v_r T^{00} + v_r P_r \right) \, ,  \qquad \\
\partial_t T^{0r} + \partial_{r}\left(v_r T^{0r}\right)
&=&  - \partial_{r} P_r  \nonumber \\
&-& \frac{1}{r} \left(v_r T^{0r} - 2r^2 \pi^{\phi \phi}
- \pi^{zz} \right) \, . \qquad \label{cons_T0r}
\eea
Due to symmetry the right-hand side of Eq.\ \eqref{cons_T0r}
has to vanish at the origin. The relaxation equations
follow from Eqs.\ (\ref{relax_bulk}), (\ref{relax_shear}),
\bea
\gamma_r \partial_t \Pi + \gamma_r v_r \partial_r \Pi
&=& \frac{1}{\tau_\pi}\left(\Pi_{NS} - \Pi \right) - I_0\, ,
\\ \nonumber
\gamma_r \partial_t \pi^{\phi \phi} +
\gamma_r v_r \partial_r \pi^{\phi \phi}
&=& \frac{1}{\tau_\pi} \left(\pi^{\phi \phi}_{NS}
- \pi^{\phi \phi} \right) \\
&-& 2 \frac{\gamma_r v_r}{r} \pi^{\phi \phi}
- I^{\phi \phi}_2 \, , \\
\gamma_r \partial_t \pi^{zz} + \gamma_r v_r \partial_r \pi^{zz}
&=& \frac{1}{\tau_\pi} \left(\pi^{zz}_{NS} - \pi^{zz} \right)
- I^{zz}_2 \, ,
\eea
where the expansion scalar is
$\theta_r = \partial_t \gamma_r + r^{-1}
\partial_r (r \gamma_r v_r)$ and
$I^{\phi\phi}_1 = I^{zz}_1 = I^{\phi\phi}_3 = I^{zz}_3 = 0$.
Note that the convective time derivative from Eq.\ (\ref{timeD})
leads to an extra term for the $\pi^{\phi \phi}$ component,
which was missed in Eq.\ (5.16) of Ref.\ \cite{Heinz:2005bw}.

The shear stress tensor components are calculated from
Eq.\ (\ref{shear}), hence the Navier-Stokes values for the bulk
viscous pressure and shear stress tensor are,
\bea
\Pi_{NS} &\equiv& - \zeta \theta_r \, ,\\
\pi^{\phi \phi}_{NS} &\equiv& 2\eta \sigma^{\phi \phi}
= \frac{2\eta}{r^2} \left( \frac{\theta_r}{3}
- \frac{\gamma_r v_r}{r} \right) \, , \\
\pi^{zz}_{NS} &\equiv& 2\eta \sigma^{zz}
= 2\eta  \frac{\theta_r}{3} \, .
\eea
Also note that the $r^{-2}$ factor in
$\pi^{\phi \phi}_{NS}$ might cause problems close to the origin.
Hence it is preferable to rewrite the relaxation equation
using the following variable:
$\tilde{\pi}^{\phi\phi} = r^{2}\pi^{\phi \phi}$.

The term $I_0$ and the relevant components of $I^{\mu \nu}_2$
are given by
\bea
I_0 &=& \frac{1}{2}\Pi \left(\theta_r
+ D \ln \frac{\beta_0}{T}\right) \, , \\
I^{\phi \phi}_2 &=&  \frac{1}{2}\pi^{\phi \phi}
\left(\theta_r + D \ln \frac{\beta_2}{T}\right) \, , \\
I^{zz}_2 &=& \frac{1}{2}\pi^{zz}\left(\theta_r +
D \ln \frac{\beta_2}{T}\right) \, .
\eea
We also have the following relations between the
cylindrically symmetric and Cartesian systems
(with similar relations between the shear stress tensor
components)
\bea
T^{0x} &=& T^{0r} \cos \phi \, ,  \\
T^{0y} &=& T^{0r} \sin \phi \, ,  \\
T^{xx} &=& T^{rr} \cos^2 \phi + r^2T^{\phi\phi}
\sin^2 \phi \, , \\
T^{yy} &=& T^{rr} \sin^2 \phi + r^2T^{\phi\phi}
\cos^2 \phi \, , \\
T^{xy} &=& \left(T^{rr} - r^2 T^{\phi \phi} \right)
\cos\phi \sin\phi \, ,
\eea
while $T^{00}$ and $T^{zz}$ remain unchanged.
The inverse transformations are
\bea
T^{0r} &=& T^{0x} \cos \phi + T^{0y} \sin \phi\, ,  \\
T^{0\phi} &=&  \left( T^{0y} \cos \phi
- T^{0x} \sin \phi \right)\!/r\, \, ,  \\
T^{rr} &=& T^{xx} \cos^2 \phi + T^{xy} \sin (2\phi)
+ T^{yy} \sin^2 \phi \, , \\
T^{\phi \phi} &=& \left[T^{xx} \sin^2 \phi
- T^{xy} \sin (2\phi) + T^{yy} \cos^2 \phi\right]\!/r^2\, ,
\qquad\\
T^{r\phi} &=& \left[\left(T^{yy} - T^{xx} \right)
\sin \phi \cos\phi  + T^{xy} \cos(2\phi)\right]\!/r \, .
\eea
These relations will be used to compare the evolution of
cylindrically symmetric and Cartesian systems.

\subsection{(2+1)--dimensional Cartesian coordinates}\label{2d}

For (2+1)--dimensional Cartesian coordinates,
the covariant derivative of any four-vector
reduces to the standard four-divergence,
since all Christoffel symbols vanish.
We assume that the system is homogeneous in the $z$
direction and the velocity, as well as
the derivative in this direction, vanish.
Hence the four-flow and four-gradient are function
of $(t, x, y)$ coordinates alone, thus
$u^{\mu} = \gamma_\perp (1,v_x,v_y,0)$,
$\partial_{\mu} = (\partial_t,\partial_x,\partial_y,0)$,
where  $\gamma_\perp = (1 - v^2_\perp)^{-1/2}$ and
$v_\perp = (v^2_x + v^2_y)^{1/2}$.

The relevant laboratory frame quantities are
\bea
N^{0} &\equiv& n \gamma_\perp \, , \label{N02d}\\
N^{x} &\equiv& N^{0} v_x \, , \label{Nx2d}\\
N^{y} &\equiv& N^{0} v_y \, , \label{Ny2d}\\
T^{00} &\equiv& (e + P)\gamma_\perp^2 - P + \pi^{00}
\, , \label{T002d}\\
T^{0x} &\equiv& (e + P )\gamma_\perp^2 v_x + \pi^{0x} \, ,
\nonumber \\
&=& v_x T^{00} + v_x P - v_x \pi^{00} + \pi^{0x}
\, , \label{T0x2d}\\
T^{0y} &\equiv& (e + P )\gamma_\perp^2 v_y + \pi^{0y} \, ,
\nonumber \\
&=& v_y T^{00} + v_y P - v_y \pi^{00} + \pi^{0y}
\, ,\label{T0y2d}\\
T^{xx} &\equiv& (e + P)\gamma_\perp^2 v^2_x + P  + \pi^{xx} \, ,
\nonumber \\ \label{Txx2d}
&=& v_x T^{0x} + P - v_x \pi^{0x} + \pi^{xx} \, , \\
T^{yy} &\equiv& (e + P)\gamma_\perp^2 v^2_y + P  + \pi^{yy} \, ,
\nonumber \\ \label{Tyy2d}
&=& v_y T^{0y} + P - v_y \pi^{0y} + \pi^{yy}\, , \\
T^{xy} &\equiv& (e + P)\gamma_\perp^2 v_x v_y + \pi^{xy} \, ,
\nonumber\\
&=& v_x T^{0y} - v_x \pi^{0y} + \pi^{xy}  \, , \nonumber \\
&=& v_y T^{0x} - v_y \pi^{0x} + \pi^{xy} \, , \label{Txy2d} \\
T^{zz} &=& P + \pi^{zz} \, .  \label{Tzz2d}
\eea
The remaining $z$-directed four-vector and tensor
components vanish, i.e., $N^{z}=0$, $T^{0z}=T^{xz}=T^{yz}=0$ and
$\pi^{0z}=\pi^{xz}=\pi^{yz}=0$.
The LRF charge density and energy density are
\bea
n \! \! &=& \! \! N^{0} \left(1 - v^2_x - v^2_y\right)^{1/2}
\, ,\\
e \! \! &=& \! \! T^{00} - \pi^{00} -
v_x (T^{0x} - \pi^{0x}) - v_y (T^{0y} - \pi^{0y}) \, , \qquad
\eea
while the velocity components from Eq.\ (\ref{velocity}) lead to
\bea
v_x &=& \label{vx2d}
\frac{T^{0x} - \pi^{0x}}{T^{00} - \pi^{00} + P} \, , \\
v_y &=& \label{vy2d}
\frac{T^{0y} - \pi^{0y}}{T^{00} - \pi^{00} + P} \, .
\eea
The velocity components can be calculated from the
equations given above using a two-dimensional root search
or via a one-dimensional root search using Eqs.\
(\ref{velo_magnitude}), (\ref{velo_components}).

Since we previously fixed and explicitly used some
shear stress tensor components in the velocity calculation,
we choose to express the remaining components in terms of the
former. The orthogonality relation (\ref{pi0i})
and the tracelessness relation (\ref{pi002}) yield
\bea\label{pi0x2d}
\pi^{0x} &=& \pi^{xx}v_x +  \pi^{xy}v_y \, , \\
\label{pi0y2d}
\pi^{0y} &=& \pi^{xy}v_x +  \pi^{yy}v_y \, , \\ \label{pi002d}
\pi^{00} &=& \pi^{xx} + \pi^{yy} + \pi^{zz} \, .
\eea
Therefore, as function of the chosen independent variables,
$\pi^{00},\pi^{0x},\pi^{0y}$,
and $\pi^{zz}$, the other shear stress tensor components are
\bea
\pi^{xx} \! \! &=& \label{pixx2d}
\! \! \left[v^2_y (\pi^{00} - \pi^{zz}) +
v_x \pi^{0x} - v_y\pi^{0y}\right]/v^2_\perp \, , \\
\pi^{yy} \! \! &=& \label{piyy2d}
\! \! \left[v^2_x (\pi^{00} - \pi^{zz}) -
v_x \pi^{0x} - v_y\pi^{0y}\right]/v^2_\perp \, , \\
\pi^{xy} \! \! &=& \label{pixy2d}
\! \! \left[-v_x v_y (\pi^{00} - \pi^{zz}) +
v_y \pi^{0x} + v_x \pi^{0y}\right]/v^2_\perp \, . \qquad
\eea
One may then check whether
the remaining orthogonality relation,
\be
\pi^{00} = \pi^{0x} v_x + \pi^{0y} v_y \, , \label{pi002d_2}
\ee
is fulfilled.
The above relations between the shear stress tensor
components become unusable in the case that the velocity in the
transverse direction approaches zero.
Therefore, in our calculations we shall neglect the above
simplifications and calculate all shear stress tensor
components explicitly.

Note that one can choose to select $\pi^{xx},\pi^{yy}$,
and $\pi^{xy}$ as independent components, therefore
$\pi^{00},\pi^{0x},\pi^{0y}$, and $\pi^{zz}$ are given
by Eqs.\ (\ref{pi0x2d})-(\ref{pi002d}) and (\ref{pi002d_2}),
see Refs.\ \cite{Chaudhuri:2007zm,Chaudhuri:2008sj}.
However, in this case the velocity iteration is two-dimensional
which may become computationally as expensive as solving
the additional transport equations.

The conservation of net charge $N^{0}$,
energy $T^{00}$, and the momentum components
$T^{0x}$ and $T^{0y}$ are
\bea
\lefteqn{\partial_t N^{0} + \partial_x (v_x N^{0})
+ \partial_y (v_y N^{0}) = 0 \, ,} \\
\lefteqn{\partial_t T^{00} + \partial_x (v_x T^{00})
+ \partial_y (v_y T^{00})  } \nonumber \\
&=& - \partial_x (v_x P - v_x \pi^{00} + \pi^{0x}) \nonumber \\
& & - \partial_y (v_y P - v_y \pi^{00} + \pi^{0y})
\, , \label{E2d}\\
\lefteqn{\partial_t T^{0x} + \partial_x (v_x T^{0x})
+ \partial_y (v_y T^{0x}) } \nonumber \\ \label{Mx2d}
&=& - \partial_x \left(P - v_x \pi^{0x} + \pi^{xx} \right)
- \partial_y \left(- v_y \pi^{0x} + \pi^{xy} \right) \, ,
\\
\lefteqn{\partial_t T^{0y} + \partial_x (v_x T^{0y})
+ \partial_y (v_y T^{0y}) }\nonumber \\ \label{My2d}
&=& - \partial_x \left(- v_x \pi^{0y} + \pi^{xy} \right)
- \partial_y \left(P - v_y \pi^{0y} + \pi^{yy} \right) \, .
\qquad
\eea
The relaxation equations for the bulk viscous pressure,
$\Pi$, and the components
$\pi^{00},\pi^{0x},\pi^{0y},\pi^{xx},\pi^{yy},\pi^{zz},
\pi^{xy}$ of the shear stress tensor are
\bea
\lefteqn{\gamma_\perp \partial_t \Pi
+ \gamma_\perp v_x \partial_x \Pi
+ \gamma_\perp v_y \partial_y \Pi}\nonumber\\
& = &
\frac{1}{\tau_\Pi}\left(\Pi_{NS} - \Pi \right) - I_0 \, ,
\label{relax_bulk2d}\\
\lefteqn{\gamma_\perp \partial_t \pi^{\mu\nu}
+ \gamma_\perp v_x \partial_x \pi^{\mu\nu}
+ \gamma_\perp v_y \partial_y \pi^{\mu\nu} } \nonumber \\
&=& \frac{1}{\tau_\pi}\left(\pi^{\mu\nu}_{NS}
- \pi^{\mu\nu} \right) -
I^{\mu \nu}_1  - I^{\mu \nu}_2 - I^{\mu \nu}_3 \, ,
\label{relax_shear2d}
\eea
where the Navier-Stokes values are
$\Pi_{NS} \equiv -\zeta \theta_\perp$ and
$\pi^{\mu\nu}_{NS} \equiv 2\eta \sigma^{\mu \nu}$,
and the expansion scalar is
$\theta_\perp = \partial_t \gamma_\perp
+ \partial_x (\gamma_\perp v_x) + \partial_y (\gamma_\perp v_y)$.

The components of the shear tensor can be calculated from
Eq.\ (\ref{shear}), which reduces to the following simple form
$\sigma^{\mu \nu} \equiv \frac{1}{2}
\left(\partial^\mu u^\nu - u^{\mu} Du^{\nu}
+ \partial^\nu u^\mu - u^{\nu}D u^{\mu}\right) -
\frac{\theta_\perp}{3} \Delta^{\mu \nu}$ in
Cartesian coordinates. Hence,
\bea
\sigma^{00} \label{s002d}
&=& \partial_t \gamma_\perp - \gamma_\perp D\gamma_\perp
+ \left(\gamma^2_\perp - 1\right)\frac{\theta_\perp}{3}  \, ,  \\
\sigma^{0x}
&=& \frac{1}{2} \left[ \partial_t (\gamma_\perp v_x) - \partial_x
\gamma_\perp \right]  \nonumber\\ \label{s0x2d}
&-& \frac{1}{2} \left[\gamma_\perp D (\gamma_\perp v_x)
+ \gamma_\perp v_x D\gamma_\perp \right]
+ \gamma_\perp^2 v_x \frac{\theta_\perp}{3} \, , \\
\sigma^{0y}
&=& \frac{1}{2}\left[\partial_t (\gamma_\perp v_y) - \partial_y
\gamma_\perp \right]  \nonumber \\ \label{s0y2d}
&-& \frac{1}{2}\left[\gamma_\perp D (\gamma_\perp v_y)
+ \gamma_\perp v_y D\gamma_\perp \right]
+ \gamma_\perp^2 v_y \frac{\theta_\perp}{3} \, , \qquad \\
\sigma^{xx}
&=& - \partial_x (\gamma_\perp v_x)
- \gamma_\perp v_x D (\gamma_\perp v_x)
\nonumber \\  \label{sxx2d}
&+& (1 + \gamma_\perp^2 v^2_x) \frac{\theta_\perp}{3} \, ,  \\
\sigma^{yy}
&=& - \partial_y (\gamma_\perp v_y)
- \gamma_\perp v_y D (\gamma_\perp v_y) \nonumber\\ \label{syy2d}
&+& (1 + \gamma_\perp^2 v^2_y) \frac{\theta_\perp}{3} \, ,  \\
\sigma^{xy}
&=& -\frac{1}{2} \left[ \partial_x (\gamma_\perp v_y)
+ \partial_y (\gamma_\perp v_x) \right]  \nonumber\\
&-& \frac{1}{2} \left[ \gamma_\perp v_x D (\gamma_\perp v_y)
+ \gamma_\perp v_y D (\gamma_\perp v_x) \right] \nonumber \\
& +&
\gamma_\perp^2 v_x v_y \frac{\theta_\perp}{3}
\, ,\label{sxy2d} \\
\sigma^{zz} &=& \frac{\theta_\perp}{3} \, ,
\eea
where
$D \equiv u^{\mu} \partial_{\mu} = \gamma_\perp \partial_t
+ \gamma_\perp v_x \partial_x + \gamma_\perp v_y \partial_y$.

The term  $I^{\mu \nu}_1 = (\pi^{\lambda\mu}u^\nu
+\pi^{\lambda\nu}u^\mu)D u_\lambda$ leads to
\bea
I^{00}_1  &=&  2 \gamma_\perp \left[\pi^{00} D \gamma_\perp
- \pi^{0x} D (\gamma_\perp v_x) \right. \nonumber\\
&-& \left. \pi^{0y} D (\gamma_\perp v_y) \right] \, , \\
I^{0x}_1 &=& \gamma_\perp \left[(\pi^{00} v_x
+ \pi^{0x})D \gamma_\perp \right.  \nonumber\\
&-& \left. (\pi^{0x} v_x  + \pi^{xx})D (\gamma_\perp v_x)
\right.  \nonumber \\
&-& \left. (\pi^{0y} v_x  + \pi^{xy})D (\gamma_\perp v_y)
\right]\, , \\
I^{0y}_1  &=&  \gamma_\perp \left[ (\pi^{00} v_y  + \pi^{0y})
D \gamma_\perp \right. \nonumber \\
&-& \left. (\pi^{0x} v_y  + \pi^{xy})
D (\gamma_\perp v_x) \right.  \nonumber \\
&-& \left. (\pi^{0y} v_y  + \pi^{yy})D (\gamma_\perp v_y)
\right] \, , \\
I^{xx}_1  &=&  2 \gamma_\perp v_x \left[ \pi^{0x}
D \gamma_\perp  - \pi^{xx}
D (\gamma_\perp v_x) \right. \nonumber \\
&-& \left. \pi^{xy} D (\gamma_\perp v_y) \right]\, , \\
I^{yy}_1 &=&  2 \gamma_\perp v_y \left[ \pi^{0y}
D \gamma_\perp  - \pi^{xy} D (\gamma_\perp v_x) \right.
\nonumber \\
&-& \left. \pi^{yy} D (\gamma_\perp v_y) \right]\, , \nonumber\\
I^{xy}_1 \! \! &=& \! \! \gamma_\perp \left[ (\pi^{0x} v_y
+ \pi^{0y}v_x)D \gamma_\perp \right. \\ \nonumber
&-& \left. (\pi^{xx} v_y  + \pi^{xy}v_x)D (\gamma_\perp v_x)
\right. \\
&-& \left. (\pi^{xy} v_y  + \pi^{yy}v_x)D (\gamma_\perp v_y)
\right]\, ,
\eea
and $I^{zz}_1 = 0$.
The terms $I_0$ and $I_2^{\mu \nu}$ are again given
by Eqs.\ (\ref{I_0}) and (\ref{I_2}), respectively.
Finally, the components of the
term $I^{\mu\nu}_3 = \pi^{\mu\lambda}
\omega^{\nu}_{\hspace*{0.1cm} \lambda} + \pi^{\nu \lambda}
\omega^{\mu}_{\hspace*{0.1cm} \lambda}$ are explicitly given by
\bea
I^{00}_3 &=& 2\left(\pi^{0x} \omega^0_{\hspace*{0.1cm} x}
+\pi^{0y} \omega^0_{\hspace*{0.1cm} y} \right) \, , \\
I^{0x}_3 &=& \pi^{00} \omega^x_{\hspace*{0.1cm} 0}
+ \pi^{0y} \omega^x_{\hspace*{0.1cm} y}
+ \pi^{xx} \omega^0_{\hspace*{0.1cm} x}
+ \pi^{xy} \omega^0_{\hspace*{0.1cm} y} \, ,\\
I^{0y}_3 &=& \pi^{00} \omega^y_{\hspace*{0.1cm} 0}
+ \pi^{0x} \omega^y_{\hspace*{0.1cm} x}
+ \pi^{xy} \omega^0_{\hspace*{0.1cm} x}
+ \pi^{yy} \omega^0_{\hspace*{0.1cm} y} \, ,\\
I^{xx}_3 &=& 2 \left( \pi^{0x} \omega^x_{\hspace*{0.1cm} 0}
+ \pi^{xy} \omega^x_{\hspace*{0.1cm} y} \right)  \, , \\
I^{yy}_3 &=& 2 \left( \pi^{0y} \omega^y_{\hspace*{0.1cm} 0}
+ \pi^{xy}
\omega^y_{\hspace*{0.1cm} x} \right) \, , \\
I^{xy}_3 &=& \pi^{0x} \omega^y_{\hspace*{0.1cm} 0}
+ \pi^{xx} \omega^y_{\hspace*{0.1cm} x}
+ \pi^{0y} \omega^x_{\hspace*{0.1cm} 0}
+ \pi^{yy} \omega^x_{\hspace*{0.1cm} y} \, ,
\qquad
\eea
and $I^{zz}_3 = 0$.
The vorticity tensor in Cartesian coordinates is given by,
$\omega^{\mu}_{\hspace*{0.1cm} \nu} \equiv \frac{1}{2}
\left(\partial_\nu u^\mu - \partial^\mu u_\nu
+ u^\mu Du_\nu - u_\nu Du^\mu \right)$,
therefore the nonvanishing vorticity tensor components are
\bea
\omega^0_{\hspace*{0.1cm} x} &=&
\frac{1}{2}\left[\partial_x \gamma_\perp
+ \partial_t (\gamma_\perp v_x) \right]
\nonumber \\ \label{w0x2d}
&+& \frac{1}{2}\left[\gamma_\perp v_x D \gamma_\perp
- \gamma_\perp D (\gamma_\perp v_x) \right]\, , \\
\omega^0_{\hspace*{0.1cm} y} &=&
\frac{1}{2}\left[\partial_y \gamma_\perp
+ \partial_t (\gamma_\perp v_y) \right]
\nonumber \\  \label{w0y2d}
&+& \frac{1}{2}\left[\gamma_\perp v_y D \gamma_\perp
- \gamma_\perp D (\gamma_\perp v_y) \right]\, , \\
\omega^x_{\hspace*{0.1cm} y} &=&
\frac{1}{2} \left[ \partial_y (\gamma_\perp v_x) - \partial_x
(\gamma_\perp v_y) \right] \nonumber  \\ \label{wxy2d}
&+& \frac{1}{2} \left[ \gamma_\perp v_y D (\gamma_\perp v_x)
- \gamma_\perp v_x D (\gamma_\perp v_y) \right] \, ,
\eea
where the vorticity tensor components satisfy the following
relations, $\omega^0_{\hspace*{0.1cm} x}
= \omega^x_{\hspace*{0.1cm} 0} = -\omega^{0x}
= \omega_{0x}$, $\omega^0_{\hspace*{0.1cm} y}
= \omega^y_{\hspace*{0.1cm} 0} = -\omega^{0y}
=\omega_{0y}$ and $\omega^x_{\hspace*{0.1cm} y}
= -\omega^y_{\hspace*{0.1cm} x} = -\omega^{xy}
= -\omega_{xy}$.

\section{Numerical methods} \label{IV}

In this section we present in detail the numerical algorithm
used to solve the equations of relativistic dissipative
fluid dynamics in (1+1)-- and (2+1)--dimensional geometries.
In our case, this will be the SHArp and Smooth Transport
Algorithm (SHASTA) \cite{BorisBook_SHASTA}.
We also briefly discuss other schemes, and conclude with
remarks on the numerical resolution and dissipative fluxes.

\subsection{One-dimensional implementation}

In (1+1)--dimensional systems the equations of
charge and energy-momentum conservation,
Eqs.\ (\ref{N01d_cons}), (\ref{T001d_cons}), (\ref{T0i1d_cons}),
are of conservation type and can be generally written as
\bea
\partial_t U + \partial_x (v_x U) = S(t,x) \, ,
\eea
where $U = U(t,x)$ is the conserved quantity, $v_x$ is
the flow velocity in $x$ direction, and $S(t,x)$ is the
source term.
The relaxation equations (\ref{relax_bulk_1d}),
(\ref{relax_shear_1d}) are of convective type.
These equations can be rearranged in conservation form
with an additional source term \cite{Muronga:2006zw,Molnar:2008fv},
\bea
\partial_t U_\pi + \partial_x (v_x U_\pi) =
U_\pi \partial_x v_x +  S_\pi(t,x) \, ,
\eea
where $U_\pi$ is $\Pi$ or $\pi$, and $S_\pi$
is the source term either from Eq.\ (\ref{relax_bulk_1d})
or Eq.\ (\ref{relax_shear_1d}) divided by
$\gamma = \left(1 - v_x^2\right)^{-1/2}$.

To solve the above type of equations numerically,
the original partial differential equations are replaced
by approximate algebraic difference equations and the
values of $U$, $v$, and $S$ are given at discrete grid points.
The conservative, or primary, variable
$U(t,x)$ is replaced by its average $U^n_i$
over the cell $i$ at coordinate point $x_i$, and
at the discrete time step $t^n$. The algorithms used in
this work belong to the class of finite-volume methods
where fluxes of the conserved quantity through the
cell boundaries are calculated or approximated.
This explicitly guarantees the conservation
of the primary variable. The velocity and source terms
are defined as a function of primary variables.
Whenever source terms contain spatial derivatives,
they are calculated by using second-order central differences,
e.g.\ $\partial_x U^{n}_{i} =  (U^{n}_{i+1} - U^{n}_{i-1})/
(2\Delta x)$.
Time derivatives in source terms are calculated using
first-order backward differences,
e.g.\ $\partial_t U^{n}_{i} =  (U^{n-1}_{i} - U^{n}_{i})/
\Delta t$.

Here we will give a brief presentation of our
numerical algorithm.
Due to its simplicity, accuracy, and easy implementation
for this study we choose the SHASTA \cite{BorisBook_SHASTA}
which was one of the first versions of
Flux Corrected Transport (FCT) algorithms in the 1970's.
Ever since, the FCT method has been extensively
tested and refined for various studies,
for example, the ETBFCT version by Boris \cite{ETBFCT},
which also forms the basis for the LCPFCT algorithm
\cite{LCPFCT}, and the YDFCT algorithm by T\'oth and Odstrcil
\cite{Toth_1996}.

These explicit higher-order monotonic numerical
methods have been especially designed to work in the presence
of strong gradients such as shocks.
Typically low-order numerical schemes result in strong
numerical diffusion due to the large truncation error,
which tends to smooth out all the structures in the solution.
Thus, low-order schemes are practically unusable unless
unrealistically small grid sizes are used.
Second-order schemes do not suffer from large numerical
diffusion, but instead from a strong numerical
dispersion, i.e., different Fourier modes propagate at
different speeds. Especially in the presence of strong
gradients like shock waves, numerical dispersion causes
unphysical ripples in the solution, which eventually
invalidates the whole calculation.

In the SHASTA this problem is solved by first
calculating a low-order solution which has a large
numerical diffusion component.
In the second step, as much diffusion as possible
is removed from the low-order solution in such a
way that no new maxima or minima are created, i.e.,
the monotonicity of the solution will be preserved.
The remaining, or residual, diffusion of numerical origin
is called numerical viscosity.
In the FCT algorithms the numerical viscosity has both
linear and non-linear contributions and
therefore must be assessed separately for the problems at hand.
This implicit numerical viscosity is, of course,
different from the well-known explicit artificial
viscosity of von Neumann, or Lax and Wendroff \cite{NumVisco2}.

The low-order, or transported and diffused, solution
in the explicit SHASTA method \cite{BorisBook_SHASTA}
is given by
\bea
\tilde{U}_i &=& \frac{1}{2} \left(Q^2_{+} \Delta_{i}
- Q^2_{-} \Delta_{i-1}\right) \nonumber \\ \label{trans_diff}
&+& \left(Q_{+} - Q_{-}\right) U^{n}_{i} + \Delta t \, S_i \, .
\eea
Here, we defined
\bea\label{diff}
\Delta_i &=& U_{i+1} - U_{i} \, , \\
Q_{\pm} &=& \frac{1/(2 \lambda) \mp v^{n}_{i}}{
1/\lambda \pm (v^{n}_{i \pm 1} - v^{n}_{i})} \, ,
\eea
where $\lambda \equiv \Delta t/\Delta x$ is the Courant
number which in the SHASTA is restricted
to values $\lambda \leq 1/2$.
The final time-advanced quantities are calculated by
subtracting the so-called antidiffusion fluxes,
$\tilde{A}$, from the transported and diffused
solution such that
\bea
U^{n+1}_i = \tilde{U}_i - \tilde{A}_i + \tilde{A}_{i-1} \, ,
\eea
where the flux-corrected antidiffusion flux is
\bea\label{antidiff}
\tilde{A}_i &=& \sigma_i \, \textrm{max}\left[0, \textrm{min}
\left(\sigma_i \tilde{\Delta}_{i+1}, |A_i|,\sigma_i
\tilde{\Delta}_{i-1} \right)\right]\, . \qquad
\eea
Here, similarly as in Eq.\ (\ref{diff}) the difference
of primary variables in adjacent cells is denoted by
$\tilde{\Delta}_i = \tilde{U}_{i+1} - \tilde{U}_i$, while
the explicit antidiffusion flux is
\bea\label{adiff1d}
A_i &=& A_{ad} \, \tilde{\Delta}_i / 8\, ,\\
\label{sig1d}
\sigma_i &=& \textrm{sgn}(A_i) \, .
\eea
In the SHASTA, $A_{ad}=1$ is the default value of the
so-called mask coefficient \cite{Book_1}.
This is a multiplicative constant which can be set to
lower values to reduce the amount of antidiffusion.

Second-order accuracy in time is obtained by applying
the SHASTA twice. First we calculate the velocity and
source terms at time step $n+1/2$.
In the second step, these half-step velocity and source
terms are used to calculate the final time-advanced
quantity $U^{n+1}_i$.
In a given cell, this can be summarized in formulas as
\bea \label{U_halfstep}
U^{n+1/2} = U^{n} \left( U^{n}, v^{n}, S^{n}, \Delta t/2,
\Delta x \right)\, , \\
U^{n+1} = U^{n} \left( U^{n}, v^{n+1/2}, S^{n+1/2}, \Delta t,
\Delta x \right) \, .
\eea
The relaxation equations are solved in a similar manner,
however in this case the source terms actually depend on
the primary variables, velocity field, and LRF quantities,
therefore their values must be saved for full and half
time steps. This requires much more memory compared to codes
which solve ideal relativistic fluid dynamics.

\subsection{Multidimensional implementation}

The (2+1)--dimensional conservation equations are
commonly written as
\bea
\partial_t U + \partial_x(v_x U) + \partial_y(v_y U)
= S(t,x,y) \, .
\eea
The cell-averaged conserved variable $U(t,x,y)$
is denoted by $U^n_{i,j}$. A standard approach to solve
such equations is to apply the dimensional or
operator splitting method, which splits the original
multidimensional equation into a sequence of
(1+1)--dimensional problems \cite{BorisBook_SHASTA_2}.

A slightly different but more efficient approach is used
in this work. We calculate the low-order transport
solution separately in the $x$ and $y$ directions by
using the (1+1)--dimensional SHASTA~\eqref{trans_diff}
without the source term.
Thus, the $x$-transported quantity $\tilde{U}^x_{i,j}$
is given as
\bea
\tilde{U}^x_{i,j} &=& \frac{1}{2}
\left[ \left( Q^{x}_{+}\right)^2 \Delta^x_{i,j}
- \left(Q^x_{-}\right)^2 \Delta^x_{i-1,j}\right]
\nonumber \\ \label{tstep_2d}
&+& \left(Q^x_{+} - Q^x_{-} \right) U^{n}_{i,j} \, , \\
Q^x_{\pm} &=& \frac{1/(2 \lambda^x) \mp (v_x)^{n}_{i,j}}{
1/\lambda^x \pm [(v_x)^{n}_{i \pm 1,j} - (v_x)^{n}_{i,j}]} \, ,
\eea
where $\Delta^x_{i,j} = U^{n}_{i+1,j} - U^{n}_{i,j}$
and $\lambda^x = \Delta x/ \Delta t \le 0.5$ is the Courant
number in the $x$ direction.
A similar formula, with $v_x$ replaced by $v_y$ and
all cell differences taken in $y$ direction, holds for the
$y$-transported quantity $\tilde{U}^y_{i,j}$.
The transported and diffused solution is then
\bea
\tilde{U}_{i, j} = \tilde{U}^x_{i,j} + \tilde{U}^y_{i,j}
- U^{n}_{i, j} + \Delta t \,S_{i, j}\,.
\eea
The advantage of this method is that it keeps the $x-y$
symmetry of the system without
the need to permute the directions in which the grid is updated.
In this case it is also possible to implement
a multidimensional flux correction in the
FCT algorithm which avoids some numerical problems and
leads to slightly smoother results compared to
the dimensional splitting method for the same mask coefficient.
To obtain second order accuracy, we use the method by
DeVore \cite{DeVore}, which is an improved
version of Zalesak's method \cite{Zalesak}.
The full solution is given by
\bea
U^{n+1}_{i,j} = \tilde{U}_{i,j} - \hat{A}^x_{i,j}
- \hat{A}^y_{i,j} + \hat{A}^x_{i-1,j}
+ \hat{A}^y_{i,j-1} \, ,
\eea
where the $\hat{A}$'s are the limited antidiffusion
fluxes given in Eqs.\ (\ref{eq:Ax_limited}) and 
(\ref{eq:Ay_limited}) below.

As in the (1+1)--dimensional case the antidiffusion
fluxes in $x$ and $y$ directions are given by
\bea\label{adiff2d}
A^{x}_{i,j} = A^x_{ad} \, \tilde{\Delta}^x_{i,j} / 8,\ \\
A^{y}_{i,j} = A^y_{ad} \, \tilde{\Delta}^y_{i,j} / 8 \, ,
\eea
where $A^x_{ad},A^y_{ad}$ are the antidiffusive mask coefficients, 
similarly to the (1+1)--dimensional case. Furthermore,
\bea
\tilde{\Delta}^x_{i,j} =  \tilde{U}_{i+1,j} - \tilde{U}_{i,j}, \\
\tilde{\Delta}^y_{i,j} =  \tilde{U}_{i,j+1} - \tilde{U}_{i,j}.
\eea
In the DeVore scheme, the antidiffusion fluxes in
$x$ and $y$ directions are first limited as in the
(1+1)--dimensional case,
\bea
\tilde{A}^x_{i,j} &=&
\sigma^x_{i,j} \, \textrm{max}\left[0, \right.
 \nonumber \\ \label{Ax}
& & \left. \textrm{min}\left(\sigma^x_{i,j}
\tilde{\Delta}^x_{i+1,j}, |A^x_{i,j}|,\sigma^x_{i,j}
\tilde{\Delta}^x_{i-1,j} \right) \right]\, , \\
\tilde{A}^y_{i,j} &=&
\sigma^y_{i,j} \, \textrm{max}\left[0, \right.
\nonumber \\ \label{Ay}
& & \left. \textrm{min}\left(\sigma^y_{i,j}
\tilde{\Delta}^y_{i,j+1}, |A^y_{i,j}|,\sigma^y_{i,j}
\tilde{\Delta}^y_{i,j-1} \right) \right]\, ,
\eea
where $\sigma^{x}_{i,j} = \textrm{sgn} (A^x_{i,j})$
and  $\sigma^{y}_{i,j} = \textrm{sgn} (A^y_{i,j})$.
Note that this additional step was introduced by DeVore into
the multidimensional flux limiting algorithm by Zalesak.

The allowed values for $U^{n+1}_{i,j}$ after the antidiffusion
stage are between
\bea
\tilde{U}^{min}_{i,j}  \! \! &=& \! \!
\textrm{min}\left(\tilde{U}_{i,j-1},\tilde{U}_{i-1,j},
\tilde{U}_{i,j},\tilde{U}_{i+1,j},\tilde{U}_{i,j+1}\right)
\, ,\qquad \\
\tilde{U}^{max}_{i,j}  \! \! &=&  \! \!
\textrm{max}\left(\tilde{U}_{i,j-1},\tilde{U}_{i-1,j},
\tilde{U}_{i,j},\tilde{U}_{i+1,j},\tilde{U}_{i,j+1}\right).
\eea
The total incoming and outgoing antidiffusive fluxes
in cell $(i, j)$ are calculated as
\bea
A^{in}_{i,j} &=& \textrm{max}\left(0,\tilde{A}^x_{i-1,j}
\right) - \textrm{min}\left(0,\tilde{A}^x_{i,j} \right)
\nonumber \\
&+& \textrm{max}\left(0,\tilde{A}^y_{i,j-1} \right)
- \textrm{min}\left(0,\tilde{A}^y_{i,j} \right) \, , \\
A^{out}_{i,j} &=& \textrm{max}\left(0,\tilde{A}^x_{i,j}
\right) - \textrm{min}\left(0,\tilde{A}^x_{i-1,j} \right)
\nonumber\\
&+& \textrm{max}\left(0,\tilde{A}^y_{i,j} \right)
- \textrm{min}\left(0,\tilde{A}^y_{i,j-1} \right) \, .
\eea
This information is then used to determine the fractions
of the incoming and outgoing fluxes,
\bea
F^{in}_{i,j} &=& \left(\tilde{U}^{max}_{i,j}
-  \tilde{U}_{i,j} \right)/A^{in}_{i,j} \, ,\\
F^{out}_{i,j} &=& \left(\tilde{U}_{i,j}
-  \tilde{U}^{min}_{i,j} \right)/A^{out}_{i,j} \, ,
\eea
which is subsequently limited so that it creates no
undershoot or overshoot in the cell it is leaving or entering.
Thus, the new antidiffusive fluxes are given as
\be
\hat{A}^x_{i,j} \!  = \! \tilde{A}^x_{i,j} \! \times \!
\left\{ \begin{array}{ll}
\! \textrm{min}\left(1,F^{in}_{i+1,j},F^{out}_{i,j}\right), &
\mbox{if \, $\tilde{A}^x_{i,j} \geq 0$} , \\
\! \textrm{min}\left(1,F^{in}_{i,j},F^{out}_{i+1,j}\right), &
\mbox{if \, $\tilde{A}^x_{i,j} < 0$},
\end{array} \right. \label{eq:Ax_limited}
\ee
and
\be
\hat{A}^y_{i,j} \!  = \! \tilde{A}^y_{i,j} \! \times \!
\left\{ \begin{array}{ll}
\! \textrm{min}\left(1,F^{in}_{i,j+1},F^{out}_{i,j}\right), &
\mbox{if \, $\tilde{A}^y_{i,j} \geq 0$} , \\
\! \textrm{min}\left(1,F^{in}_{i,j},F^{out}_{i,j+1}\right), &
\mbox{if \, $\tilde{A}^y_{i,j} < 0$}.
\end{array} \right.\label{eq:Ay_limited}
\ee
This (2+1)--dimensional numerical scheme can be generalized
to (3+1) dimensions by extending the method
to another spatial direction and repeating the above steps.

\subsection{Other numerical schemes}

Computational fluid dynamics (CDF) is a constantly growing
field of research. There is a vast amount of methods which
have been designed to solve the special relativistic
fluid-dynamical equations in the perfect fluid limit,
see Ref.\ \cite{Marti_2003} and references therein.

In applications to relativistic heavy-ion collision the
FCT-SHASTA and RHLLE methods have been systematically
explored for various
test problems and shown to give excellent agreement
\cite{Schneider:1993gd,Rischke:1995ir,Rischke:1998fq}.
This is one of the reasons why we have chosen the SHASTA
for our study. There are other well-known methods
widely used in astrophysics and heavy-ion physics,
such as Smoothed-Particle Hydrodynamics (SPH) \cite{SPH}
which has been recently extended to dissipative fluids
\cite{SPH_dissipative}, or the Particle-In-Cell (PIC) method
\cite{PIC}, but since they are completely different
from finite-volume schemes we will not go into details.

Other methods of interest such as High-Resolution
Shock-Capturing (HRSC) methods based on the exact
or approximate Riemann solution proved to be superior
to SHASTA \cite{Schneider:1993gd,Marti_2003,Font_2008}.
However, in case of dissipative fluids such methods become
difficult to apply due to the fact that there are no
known analytic or approximate solutions for the Riemann problem.
Recently, new methods have been developed to solve the
hyperbolic equations of conservation or relaxation type
which sidestep the need of Riemann solvers but have an
accuracy comparable to HRSC schemes.
These are new High-Resolution Central Schemes (HRCS)
improving on the Lax-Friedrichs central scheme \cite{Lax}.
The most important of these are the Nessayu-Tadmor
(NT) \cite{NT_scheme} and the Kurganov-Tadmor (KT) \cite{KT_scheme}
schemes, see Ref.\ \cite{cenpack} for a collection of references.

The KT scheme improves upon the NT scheme using information
about the local propagation of speeds, which becomes
problematic to evaluate for the IS equations.
However, it gives excellent results for perfect fluids
\cite{LucasSerrano:2004aq}. An important extension of the
NT scheme was made by Pareschi \cite{Pareschi} to describe
both the stiff and unstiff regions of hyperbolic
relaxation equations such as the IS equations or equations
of \"Ottinger-Grmela type \cite{Grmela:1997zz}.
In the latter case, this method has been shown to
provide robust results and excellent agreement between
the (1+1)-- and (2+1)--dimensional cases \cite{Dusling:2007gi}.
Following this work we also made use of both the NT
and KT schemes and compared them with SHASTA for
the (1+1)--dimensional evolution of a perfect fluid.
The results are very robust and agree very well.
Therefore, without much more efficient methods at hand
we simply choose to solve the IS equations with the SHASTA.

\subsection{Remarks on numerical resolution}

Fluid dynamics is a theory which is valid on time and length
scales which are larger than the underlying microscopic time
and length scales.
In solving the equations of fluid dynamics numerically, we 
should be able to resolve all relevant time and length scales 
in the problem. In practice this means that the grid spacing 
$\Delta x$ and time step $\Delta t$ should be smaller than 
any of these scales. In perfect fluid dynamics, or 
in the Navier-Stokes theory, all scales are macroscopic, i.e.,
they are inversely proportional to the gradients of the 
fluid-dynamical variables like flow field and densities. Thus 
it is sufficient to have a numerical resolution that correctly 
resolves the macroscopic structures.

In the IS theory we also need to solve the relaxation 
equations for the dissipative currents. In this case the relevant
time scale to be resolved is the relaxation time 
$\tau_R$, which is of the order of the mean time between the 
collision of particles. Thus, the time step should be
chosen such that $\Delta t \ll \tau_R$. 
If $\tau_R$ is much smaller than the macroscopic scales, this
might require very high resolution and
therefore lead to very demanding calculations. 
However, in modeling heavy ion collisions, an application 
which we mainly have in mind, scale separation
by several orders of magnitude is not expected throughout 
the whole fluid-dynamical evolution. 

There exist specialized methods \cite{Pareschi} to 
solve the equations in stiff regions. However, we do not consider
these methods here, but simply choose sufficiently high resolution
to resolve both the macroscopic and relaxation time scales.
Therefore, we solve simultaneously both the conservation 
and the relaxation equations with the same numerical resolution
and scheme.

\subsection{Remarks on dissipative fluxes}

In relativistic dissipative fluid dynamics,
the components of $T^{\mu \nu}$ and $\pi^{\mu \nu}$
cannot take arbitrary values.
Obvious physical constraints are
that the LRF energy density must be positive semi-definite
and the velocity must be bounded from above by
the speed of light, i.e., $e \geq 0$ and $v \leq 1$.
Another constraint follows from the equation for
energy conservation,
\begin{equation}
\partial_\mu T^{0\mu} =
\partial_t T^{00} + \nabla \cdot (\vec{\tilde{v}} T^{00})
=0\;,
\end{equation}
where $\tilde{v}^i \equiv T^{0i}/T^{00}$. In order to
have causal propagation of energy, we have to require that
$|\vec{\tilde{v}}| \leq 1$, i.e.,
\begin{equation}
\sqrt{- T^{0i} T_{0i}} \leq T^{00}\,.
\label{eq:T0i_condition}
\end{equation}
For perfect fluids, because of Eq.\ \eqref{eq:v_1d},
this condition guarantees both $e \geq 0$ and 
$v\leq 1$, provided that the pressure is positive.
However, in dissipative fluid dynamics this is not
necessarily true, since the condition
\eqref{eq:T0i_condition} is sufficient
only if the effective pressure is positive. For example,
neglecting the shear pressure this leads
to the condition $\Pi > -p$ for the bulk viscous pressure.

The IS theory does not itself restrict the values
of the dissipative quantities. In principle any value of
shear and bulk pressure can be used, e.g.\ as an
initial condition. However, the applicability of the
theory requires that the dissipative currents give sufficiently
small corrections to the equilibrium quantities.
For the shear and bulk pressure this requirement can
be stated as
\bea
\label{eq:Pi_condition}
|\Pi| &<& C p,	\\
\label{eq:pi_condition}
|\pi^{\mu\nu}| &<& C |T^{\mu\nu}_{eq}|,
\eea
where $C$ is a constant of order, but smaller than, one.
If these conditions are not satisfied, fluid dynamics
is not expected to give a reasonable description of the
space-time evolution of the system
and the numerical calculation can become unstable \cite{Molnar:2008fv}.
To protect the code from these numerical
instabilities the conditions \eqref{eq:Pi_condition} and
\eqref{eq:pi_condition} are always enforced.
This means that after each time step the above
conditions are checked and $\pi^{\mu\nu}$ and $\Pi$
are adjusted accordingly. 
We note that the above conditions may be enforced 
before the velocity root search, in which case we have
to compare to the values of $T^{\mu\nu}_{eq}$ and $p$ 
from the previous time step. Alternatively, one can apply
these limiters inside the root search algorithm. In this
case the limiters are applied simultaneously with solving for the
LRF densities and the velocity. In every iteration of the velocity 
root search the shear and bulk viscous pressure are compared 
to the values of $T^{\mu\nu}_{eq}$ and $p$ at the current time 
level. This guarantees that the conditions are always fulfilled, 
but the drawback is that this is computationally more 
expensive.
In situations where we expect fluid dynamics to give
a reasonable description these conditions
need not be enforced. However, if they are violated only
in small regions of space-time, i.e., few cells
or few time steps, enforcing the inequalities can
prevent these regions to invalidate the whole
calculation. Naturally, if the inequalities are
violated in large regions of space-time,
it signals that fluid dynamics is no longer a
valid theory for such situations.

\section{Results of comparisons} \label{V}

In this section we apply the different numerical
schemes described above to the relativistic Riemann
problem in (1+1) and (2+1) dimensions.
In (1+1) dimensions the Riemann problem
is analytically solvable for perfect fluids.
Thus, it provides an important test case to compare
the performance and accuracy of different numerical algorithms.

Unfortunately, analytic solutions for the one-dimensional
viscous Riemann problem are, to the best of our knowledge,
not known. However, this type of one-dimensional test
was performed previously: our fluid-dynamical
calculations with non-zero viscosity were shown to give
good agreement with kinetic theory simulations using the
Boltzmann Approach to MultiParton Scatterings (BAMPS)
\cite{Xu:2004mz} parton cascade code
\cite{Bouras:2009zz,Bouras:2009nn}.
The purpose of our tests here are to show that a more
complex (2+1)--dimensional code can, with similar initial
conditions, remarkably well reproduce our earlier
results for (1+1) dimensions. This confirms that the
numerical method produces correct answers in these test
scenarios, and gives us confidence that it can be
successfully used to study phenomena where
dissipation plays an important role.

We shall proceed as follows: First, the Riemann problem
is briefly introduced and its analytic solution in (1+1)
dimensions is compared with numerical solutions in the
perfect fluid limit. Here we compare the SHASTA, the NT, and
the KT numerical schemes. They all give comparable results
and can reproduce the analytic results with sufficiently
good numerical resolution. This gives confidence that any
of the schemes forms a good basis to extend the calculation
to multidimensional problems as well as to non-zero viscosity.
In this work these extensions are made by using the SHASTA.

Second, the numerical solutions for the (1+1)--dimensional
Riemann problem with non-zero shear viscosity are shown.
We compare results from the (2+1)--dimensional code to the
results from the (1+1)--dimensional code and show that
both codes yield, to good accuracy, the same results.

Finally, the numerical solutions of the (2+1)--dimensional,
azimuthally symmetric Riemann problem with non-zero
shear viscosity are studied. We compare the results from the
(1+1)--dimensional azimuthally symmetric code to the results
from the (2+1)--dimensional code. Again, these calculations
are in excellent agreement with each other.

\subsection{The Riemann problem}

The initial setup for the (1+1)--dimensional Riemann problem
consists of two states with constant pressure,
$p_0$ and $p_4$, separated by a membrane at $z=0$.
The matter is initially at rest on both sides and homogeneous
in the transverse directions. After the membrane is removed,
in thermodynamically normal matter~\cite{Rischke:1995ir} 
there is a shock wave traveling into the region with
lower pressure, and a rarefaction fan into the region
with larger pressure. The interface between the two regions
moves at a constant velocity and is called the shock plateau.
In dissipative fluids due to non-zero viscosity the initial 
sharp discontinuity will be smeared out and the 
quantities will change smoothly rather than discontinuously.

In numerical calculations, unless stated otherwise,
we have fixed the parameters as follows:
The local Courant number is $\lambda^x = \lambda^y = 0.4$,
and the comparison is made at $t = 4$ fm.
The cell sizes $\Delta x, \Delta y$, and
the antidiffusion coefficients, $A^{x}_{ad},A^{y}_{ad}$
are specified separately in all cases.

The energy density in local equilibrium is given by
$e=\frac{3g}{\pi^2} T^4$ where $g=16$ is the number of
degrees of freedom. Therefore, on the left and right-hand
side of the initial discontinuity the energy densities
correspond to the following temperatures: on the left
$T_0=0.4$ GeV and on the right $T_4=0.2$ GeV.
The bulk viscosity to entropy density ratio $\zeta/s$
and the shear viscosity to entropy density ratio $\eta/s$ are
taken as a constant, where the entropy density $s=s_{eq}$
is fixed to its equilibrium value, $s_{eq} = \frac{4g}{\pi^2} T^3$.
In all test cases we start from local thermal equilibrium, 
i.e., initially $\pi^{\mu \nu} = 0$.
We show only results with shear viscosity, but we have tested 
that we get similar results with non-zero bulk viscous pressure.

\subsection{Comparing different methods in perfect fluid dynamics}

\begin{figure*}[!ht]
\centering
\includegraphics[width = 18cm]{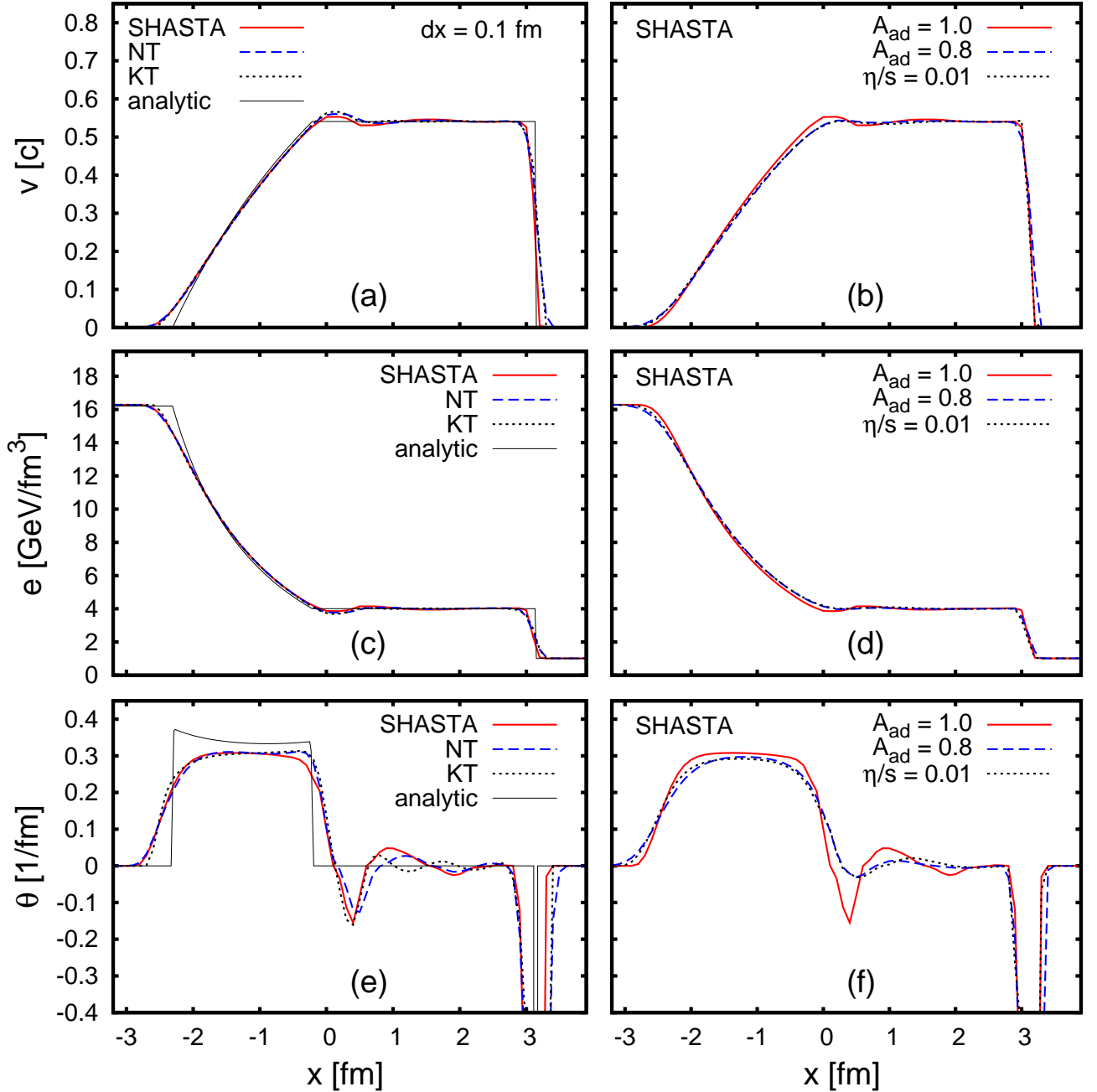}
\caption{(Color online) The analytic (thin line) and numerical 
solutions of the relativistic Riemann problem on a grid with
$N_x=100$ cells with $\Delta x=0.1$ fm, after $N_t=100$
time steps at $t=4$ fm/c.
(a) the collective flow velocity of matter, $v$,
calculated with the SHASTA (continuous line), and the
NT (dashed line) and KT (dotted line) algorithms.
(b) The velocity, $v$, calculated with SHASTA
using a mask coefficient $A_{ad} = 1.0$ (continuous line),
$A_{ad} = 0.8$ (dashed line), and vSHASTA with
$A_{ad}=1.0$ and $\eta/s=0.01$ (dotted line).
Similarly, the LRF energy density, $e$, and the
invariant expansion rate, $\theta$, are shown in the panels
(c), (d), and (e), (f), respectively.}
\label{fig_01}
\end{figure*}

\begin{figure*}[!ht]
\centering
\includegraphics[width = 18cm]{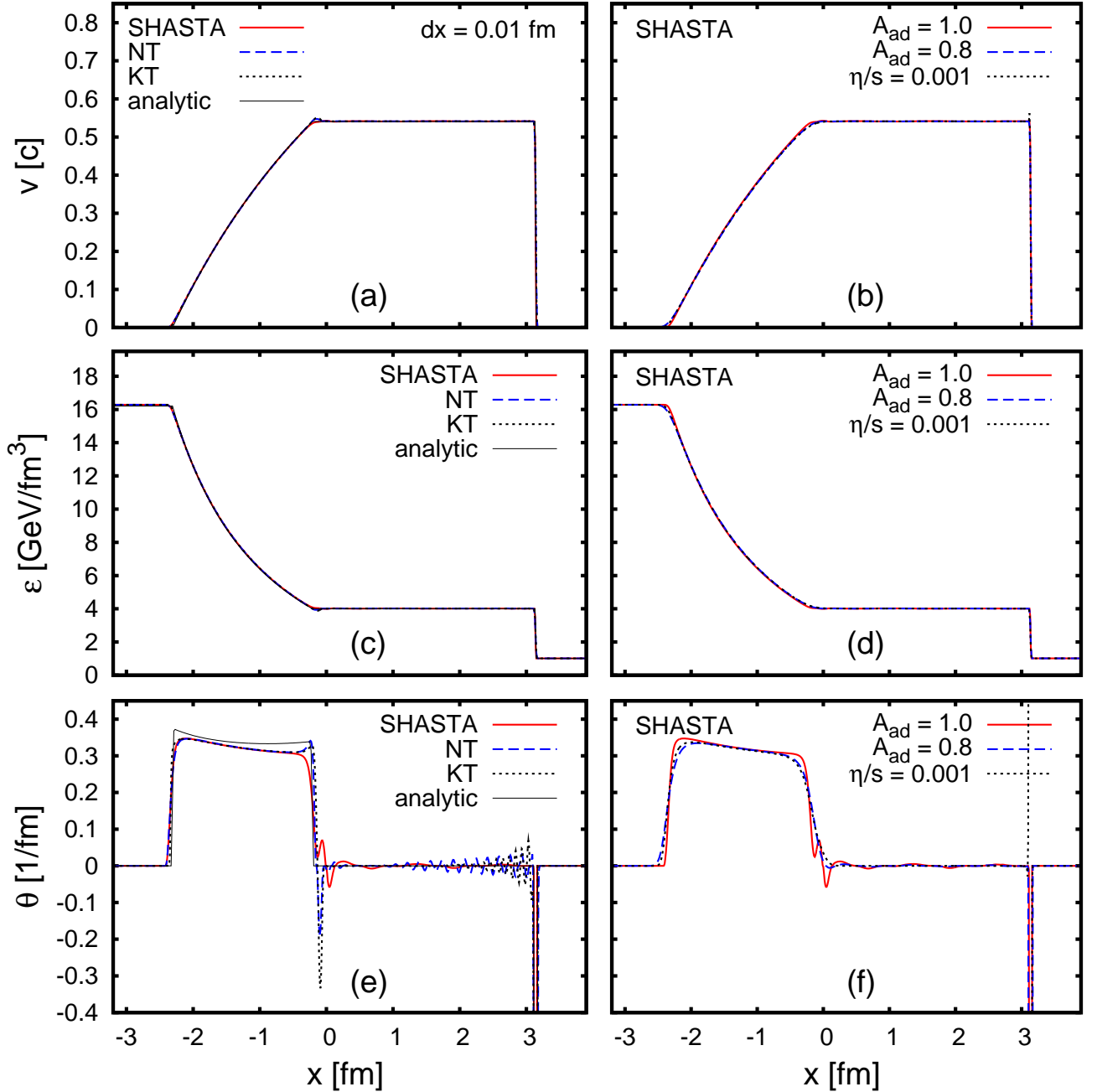}
\caption{(Color online) As in Fig.\ \ref{fig_01}, except for
$N_x=1000$ cells with $\Delta x=0.01$ fm, after $N_t=1000$
time steps at $t=4$ fm/c. Analogously, the shear viscosity to
entropy density ratio in the vSHASTA calculations
shown in panels (b), (d), (f) is $\eta/s=0.001$.}
\label{fig_02}
\end{figure*}

\begin{figure*}[!ht]
\centering
\includegraphics[width=18cm]{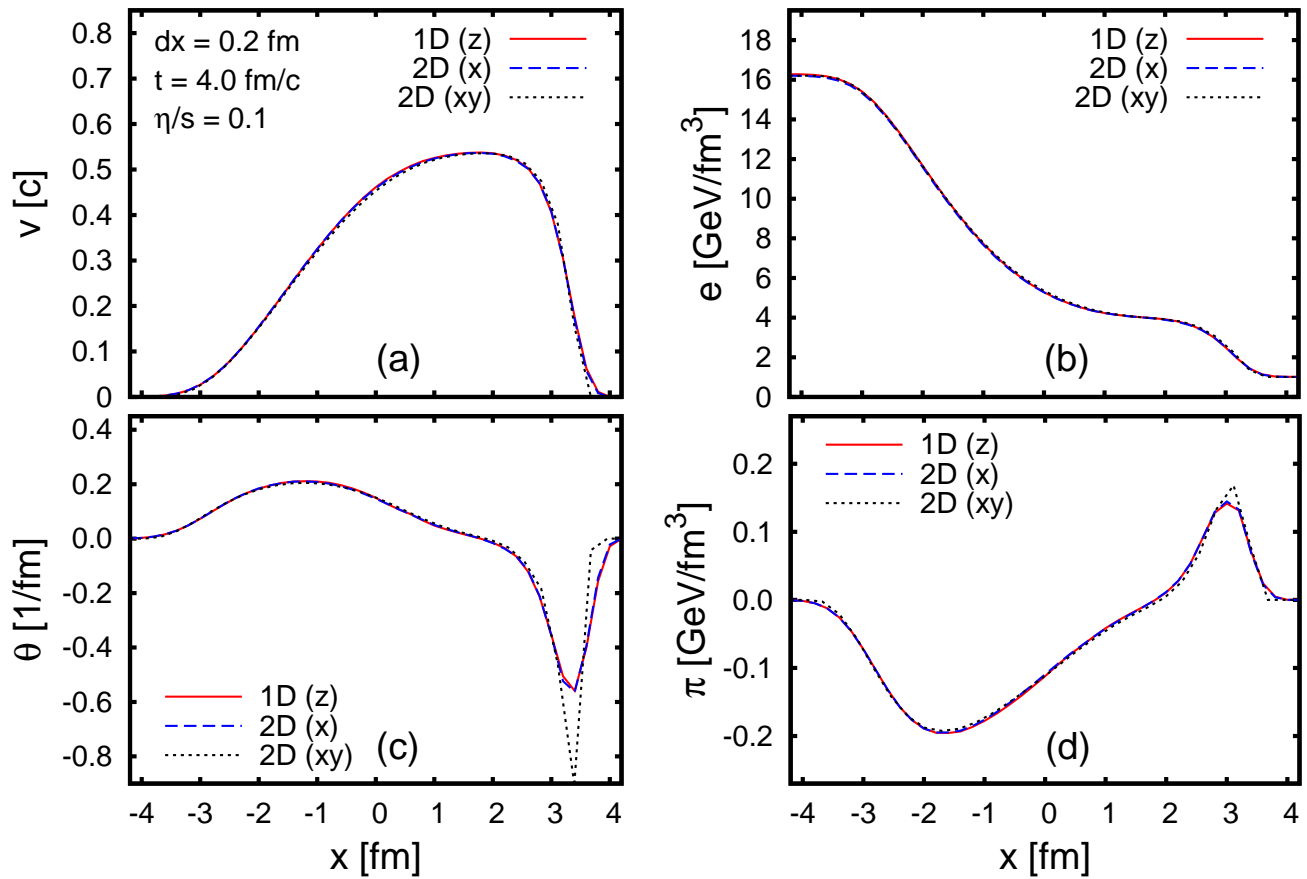}
\caption{(Color online) The numerical solution of the relativistic
Riemann problem with $\eta/s = 0.1$ on a symmetric grid
with $N_x=N_y=200$ cells with $\Delta x=0.2$ fm,
after $N_t=50$ time-steps at $t=4$ fm/c.
In all figures, the full line shows the one-dimensional
evolution of matter. The dashed line shows the two-dimensional
solution in $x$ direction, while the dotted line shows the
solution in the diagonal $x=y$ direction.
(a) The collective flow velocity of matter, $v$,
(b) the LRF energy density, $e$,
(c) the invariant expansion rate $\theta$, and
(d) the shear viscous pressure, $\pi$.
}
\label{fig_03}
\end{figure*}

\begin{figure*}[!ht]
\centering
\includegraphics[width=18cm]{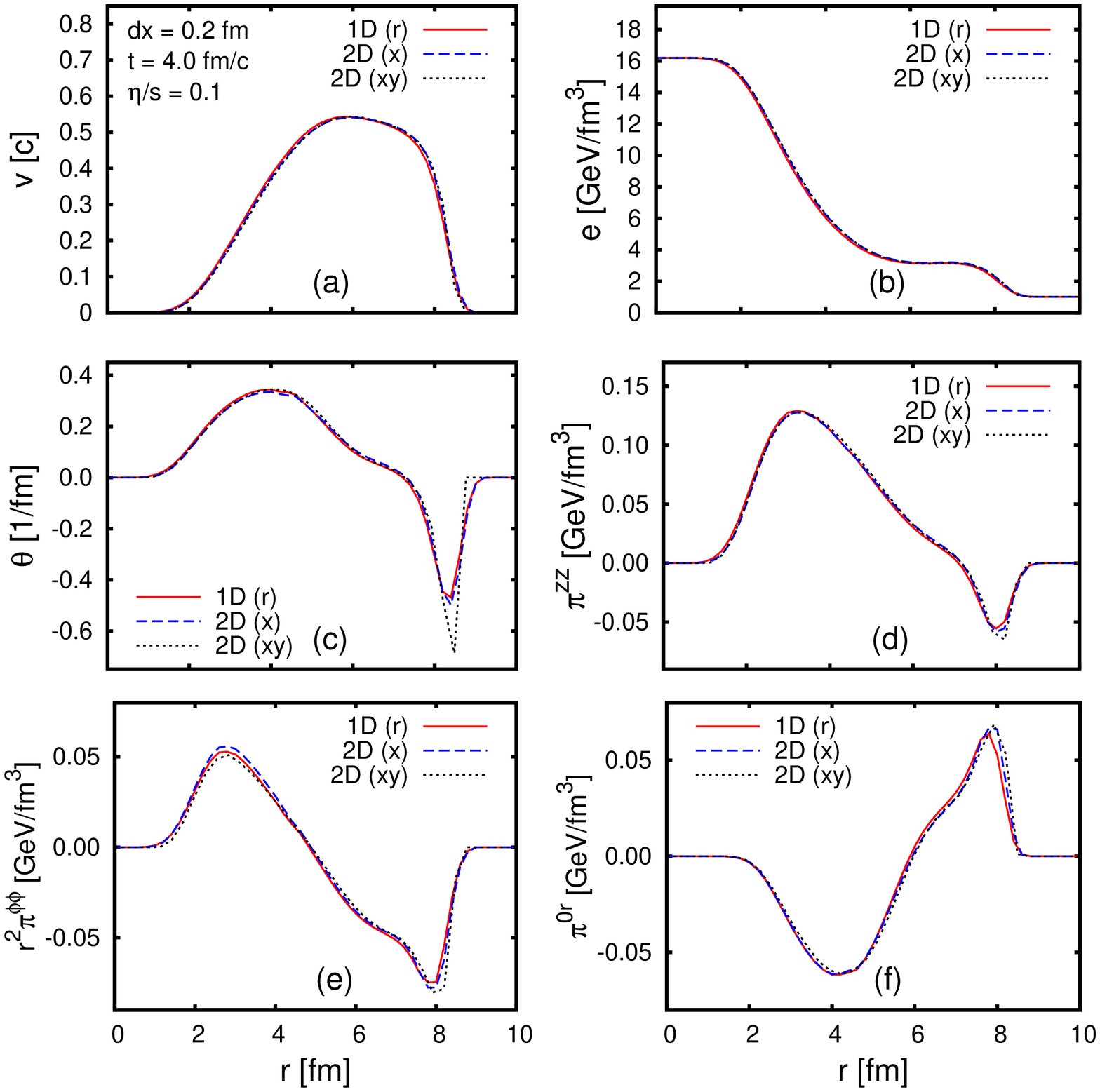}
\caption{(Color online) The numerical solution of the relativistic
Riemann problem in cylindrical geometry, with $\eta/s = 0.1$
on a symmetric grid with $N_x=N_y=200$ cells
with $\Delta x=0.2$ fm, after $N_t=50$ time-steps at $t=4$ fm/c.
In all figures, the full line shows the one-dimensional
evolution of matter. The dashed line shows the two-dimensional
solution in the $x$ direction, while the dotted line shows
the solution in the diagonal $x=y$ direction.
(a) The collective flow velocity of matter, $v$,
(b) the LRF energy density, $e$, and
(c) the invariant expansion rate, $\theta$.
The shear stress components $\pi^{zz}$,
$r^2 \pi^{\phi\phi}$, and $\pi^{0r}$ are shown in panels
(d), (e), and (f), respectively. }
\label{fig_04}
\end{figure*}

The first test compares how well the different numerical
methods can reproduce the analytic Riemann solution
\cite{Schneider:1993gd}
in the perfect-fluid limit. The left panel of Fig.\
\ref{fig_01} shows the velocity $v$,
the LRF energy density $e$, and the expansion rate
$\theta$ calculated with the SHASTA, the NT, and the KT schemes
compared with the analytic solution. The numerical
calculations in the figure are made with cell size
$\Delta x = 0.1$ fm and $\Delta t =0.04$ fm$/$c.
We used the non-staggered version of the HRCS schemes with 
a minmod limiter ($\theta = 2$) which ensures that no local 
extrema are introduced, see Eq.\ (4.9) in Ref. \cite{KT_scheme}.

All algorithms reproduce the analytic solution with
nearly the same accuracy and numerical artefacts.
In particular, all methods show long-wavelength oscillations
which are best visible in the expansion rate,
in the region between the rarefaction tail and the shock wave.
The HRCS calculations show
a somewhat larger overshoot for the velocity at the
contact discontinuity as well as a more diffused
shock front compared to SHASTA with $A_{ad}=1.0$.

We also compared the above SHASTA result with a calculation
with a reduced mask coefficient $A_{ad}=0.8$, shown in
the right panel of Fig.\ \ref{fig_01}.
This reduction strongly suppressed the unphysical
oscillations in the numerical solution, but leads also
to more diffusive profiles. Furthermore, with the
standard mask coefficient we have used the viscous
SHASTA (vSHASTA)\footnote{
Our abbreviation only specifies that next to the
conservation equations we also solve the relaxation equations
of the physical viscosity using SHASTA.}
solver with a small physical viscosity
$\eta/s = 0.01$ and $A_{ad}=1.0$. This very closely reproduces
the $A_{ad}=0.8$ results with $\eta/s = 0$, especially
at the smooth parts of the solution.
Therefore, albeit small discrepancies exist at the shock font, 
we can conclude that our numerical solutions
with the reduced antidiffusion mask coefficient have an
additional numerical viscosity corresponding to
$\eta/s \approx 0.01$ compared to the $A_{ad} = 1.0$ case.

Since all numerical calculations only approximate the
exact solution, there is always some residual numerical
viscosity in the solution. In fact, some amount of
numerical viscosity is required to stabilize
the solution. However, this residual numerical viscosity
can be made arbitrary small by increasing
the resolution. This is demonstrated in
Figs.\ \ref{fig_02}(a), (c), (e), where all numerical
algorithms considered reproduce the analytic solution
almost perfectly with a cell size of $\Delta x = 0.01$ fm and
$\Delta t = 0.004$ fm$/$c. Also, the additional numerical
viscosity resulting from the reduction of the mask
coefficient $A_{ad}$ scales approximately with the
cell size for a constant Courant number.
This is demonstrated in the right panel of Fig.\ \ref{fig_02},
where we found that the additional numerical viscosity
corresponds to $\eta/s \approx 0.001$.
We have checked that we get similar results also with
other initial temperature ratios.

\subsection{Comparison between the one-
and two-dimensional solutions}

The next numerical tests consist of comparing the
(1+1)--dimensional solution to the (2+1)--dimensional
solution of the one-dimensional Riemann problem in
Cartesian coordinates. The one-dimensional Riemann
problem can be initialized on a two-dimensional grid in
several different ways. We study here two different
initializations. In the first case, the
initial discontinuity is along the $y$ axis, i.e., on
the $x=0$ plane. In the second case
we place the discontinuity on the $y = -x$ plane.
These two cases are compared to the (1+1)--dimensional
calculation. Here both one- and two-dimensional
calculations are done using the vSHASTA algorithm,
with $A_{ad} = 0.8$, grid size $\Delta x = 0.2$ fm and
non-zero shear viscosity $\eta/s = 0.1$ in all cases.

In the simple one-dimensional formulation, there are only two
dissipative quantities to propagate, $\Pi_1$ and $\pi_1$,
while the other shear stress tensor components are
straightforward to calculate.
In the two-dimensional setup we always propagate all
non-vanishing dissipative tensor components,
$\Pi_2,\pi^{00}_2,\pi^{0x}_2,\pi^{0y}_2,\pi^{xx}_2,
\pi^{yy}_2,\pi^{xy}_2,\pi^{zz}_2$.
Because there is only one independent shear stress
component $\pi$ in the one-dimensional Riemann problem, any
of the non-vanishing shear stress components in the two-dimensional
calculations can be used to extract $\pi$. The simplest 
possibility is to
use $\pi = -2\pi^{zz}$, because $\pi^{zz}$ is independent
of the orientation of the initial state in the $(x,y)$-plane.

The result of the comparison between the one- and the
two-dimensional calculations is shown in Fig.\ \ref{fig_03},
where we compare the velocity $v$, the LRF energy density $e$,
the expansion rate $\theta$, and the shear pressure $\pi$.
The velocity $v=v_z$ in the one-dimensional calculation,
while $v=v_x$ or $v=\sqrt{v^2_x + v^2_y}$ in the
two-dimensional cases. In the two-dimensional calculations
the quantities are plotted along the $x$ axis when the
initial discontinuity is at $x=0$, or along the $y=x$
line when the discontinuity is in the $y = -x$ plane.

When the initial discontinuity is in the $x=0$ plane
the two-dimensional SHASTA reduces essentially to
the one-dimensional one. This is because there are no
gradients in the $y$ direction and $v_y=0$.
Therefore, in this case we expect very good agreement
between the one- and two-dimensional calculations.
This is confirmed in Fig.\ \ref{fig_03}, where the
two-dimensional calculation (dashed line) is basically on
top of the one-dimensional calculation (solid line).

When the initial discontinuity is along the $y = -x$ plane,
there are gradients in both $x$ and $y$ directions and
both velocity components $v_x$ and $v_y$ are non-zero.
This calculation is shown as dotted line in
Fig.\ \ref{fig_03}. The agreement with the
(1+1)--dimensional results is still very good,
although the two-dimensional algorithm gives somewhat
sharper profiles in the shock region.

The next test compares the (1+1)--dimensional solution
in cylindrically symmetric coordinates from Sec.\ \ref{1d_r}
against the two-dimensional solution in Cartesian coordinates
with cylindrically symmetric initial conditions.
This tests how well the two-dimensional system
keeps its symmetry in time
and performs compared to the one-dimensional counterpart.

The initial discontinuity lies on a circle with
radius $r_0 = 5$ fm, with a cell size of $\Delta r = 0.2$ fm
in both cases. The velocity and position in the
one-dimensional case is $v=v_r$ and $x=r$,
while in the two-dimensional case
$v=\sqrt{v^2_x + v^2_y}$ and $r=\sqrt{x^2 + y^2}$.
The first two-dimensional result compares the
evolution of the system along the $x$ axis, i.e., $y=0$,
while the second one does this along the diagonal, $x=y$.
These are plotted with dashed and dotted lines, respectively,
against the one-dimensional solution (solid line) in
Fig.\ \ref{fig_04}. The other plots show the expansion rate,
and the shear stress tensor components,
$\pi^{zz},r^2 \pi^{\phi \phi}$, and $\pi^{0r}$ as
calculated from the different equations in Secs.\
\ref{1d_r} and \ref{2d}.

Similarly as before, the results are nearly the same,
however, differences in the diagonal direction are visible
and more pronounced than along the coordinate axis,
due to the finite resolution. The agreement will obviously
get better by decreasing the cell size and time step.

\section{Conclusions} \label{VI}

In this paper, we have studied numerical algorithms to
solve the IS theory for relativistic dissipative fluid
dynamics. First, we briefly reviewed the IS theory
and wrote the IS equations for (1+1)-- and (2+1)--dimensional
systems in Cartesian coordinates, and for (1+1)--dimensional
azimuthally symmetric systems in cylindrical coordinates.
For the sake of completeness we present the (3+1)--dimensional
equations in Cartesian coordinates and the (2+1)--dimensional
boost-invariant and (3+1)--dimensional equations in 
$(\tau, x, y, \eta)$ coordinates in the Appendices. We also gave a
detailed introduction to the FCT-SHASTA method for one--
and multidimensional applications, together with a
brief discussion on the HRCS methods NT and KT.
We also discussed relationship between microscopic and 
macroscopic scales, as well as physical limitations for the
components of the energy-momentum tensor.

In our first numerical comparison we solved the
(1+1)--dimensional Riemann problem in the perfect-fluid limit.
This problem has an analytic solution which allowed
us to make a definite comparison of performance
and accuracy of the different numerical algorithms.
For this problem all the algorithms considered here,
i.e., the  NT, KT, and SHASTA methods,
gave very similar results. All of them could reproduce
the analytic solution with a very high
precision with sufficiently high numerical resolution. Moreover,
with the same resolution the accuracy of the methods was found to
be similar, i.e., none of them showed significantly faster
convergence to the analytic solution when the grid spacing was
decreased. For this reason we have chosen
SHASTA for all the other geometries as well as
for all calculations with non-zero viscosity.

We further studied the effect of the mask coefficient
$A_{ad}$ in the SHASTA. This numerical parameter
controls the amount of numerical diffusion in the algorithm.
It was found that a reduction of the coefficient by 20\%
from the default value smoothens unphysical
sharp structures in the solution, especially in the expansion
rate, and at the same time only increases the
numerical viscosity by a small amount.

In the case of non-zero viscosity, the analytic solution
to the Riemann problem is not known. However, we have
demonstrated earlier that our (1+1)--dimensional code
is in good agreement with kinetic-theory 
calculations~\cite{Bouras:2009zz,Bouras:2009nn}.
In this work we have first applied both (1+1)-- and
(2+1)--dimensional Cartesian implementations of the code
to the same (1+1)--dimensional Riemann problem.
In this case we have chosen a non-zero shear viscosity,
$\eta/s = 0.1$.
If the discontinuity in the initial energy-density
profile was chosen to be along one of the coordinate axes,
perfect agreement between the one- and the
two-dimensional codes was found. In the case where the
initial discontinuity was chosen to be along the $y = -x$ plane,
a slight difference between the two codes near the shock front
was found. The results were shown for a rather large grid
spacing $\Delta x = 0.2$ fm; the agreement was found to
improve significantly for smaller grid spacing.
A similar comparison between the (1+1)--dimensional solution in
cylindrical coordinates versus the (2+1)--dimensional
solution in Cartesian coordinates with cylindrically
symmetric initial condition confirmed that our method works
well also for problems in more than one spatial dimension.

In this work, we have demonstrated the applicability
of FCT-SHASTA to solve the conservation equations of
causal relativistic dissipative fluid dynamics simultaneously 
with relaxation transport equations. In the future, 
we intend to extend this method to full (3+1)--dimensional geometries.
We plan a detailed comparison with calculations done in the
framework of kinetic theory \cite{Bouras_NEW}, as well as studies of
collective flow in relativistic heavy-ion collisions.


\section*{ACKNOWLEDGEMENTS}
The authors would like to thank L. P. Csernai and A. Dumitru
for useful discussions, and P. Huovinen for reading the
manuscript and valuable comments.

E.\ Moln\'ar gratefully acknowledges partial support by
the Alexander von Humboldt foundation.
H.\ Niemi was supported by the Extreme Matter Institute (EMMI).
This work was supported by the Helmholtz International Center
for FAIR within the framework of the LOEWE program launched
by the State of Hesse.



\appendix
\section{(2+1)--dimensional boost-invariant expansion}
\label{2d_lc}


Because it is very important for modeling ultrarelativistic
heavy-ion collisions, we discuss the (2+1)--dimensional
boost-invariant equations of motion.
The metric tensors are
$g^{\mu \nu} = \textrm{diag}(1,-1,-1,-1/\tau^2)$ and
$g_{\mu \nu} = \textrm{diag}(1,-1,-1,-\tau^2)$, leading to
$g = \tau^2$, where $\tau = (t^2 - z^2)^{-1/2}$
is the longitudinal proper time and
$\eta = 1/2 \ln \left[(t + z)/(t - z)\right]$
is the space-time rapidity
(which is not to be confused with the shear viscosity
coefficient).
The only nonvanishing Christoffel symbols are
$\Gamma^{\eta}_{\eta \tau} = \Gamma^{\eta}_{\tau \eta}
= \tau^{-1}$ and $\Gamma^{\tau}_{\eta \eta} = \tau$.

The equations of relativistic dissipative fluid dynamics
can be easily derived from the results
in Cartesian coordinates, cf.\ Sec.\ \ref{2d}.
In order to obtain the equations for the boost-invariant
case, the indices for time $t$ and spatial $z$ direction have
to be replaced by $\tau$ and $\eta$ in the four-vector and
tensor components,
$(0,x,y,z) \rightarrow (\tau,x,y,\eta)$.
Therefore, we easily find that all laboratory
frame quantities can be written in the same way as in
Eqs.\ (\ref{N02d}) - (\ref{Txy2d}),
with the exception of the term in Eq.\ (\ref{Tzz2d}),
which becomes, $T^{zz} \rightarrow T^{\eta\eta}
\equiv P/\tau^2 + \pi^{\eta \eta}$.
This means that $N^\eta=0$,
$T^{\tau \eta}=T^{x\eta}=T^{y\eta}=0$, and
$\pi^{\tau \eta} = \pi^{x\eta} = \pi^{y\eta}=0$.

The LRF charge density, energy density, and velocity
are calculated the same way as in Sec.\ \ref{2d}.
The charge conservation equation is
\begin{equation}
\partial_\tau N^{\tau}
+ \partial_x (v_x N^{\tau}) + \partial_y (v_y N^{\tau})
= - \frac{1}{\tau} N^{\tau} \, .
\end{equation}
The energy conservation equation follows from
$\frac{1}{\sqrt{g}}\partial_{\mu}
\left(\sqrt{g} \, T^{\mu \tau} \right)
+ \Gamma^{\tau}_{\mu \beta} T^{\mu \beta} = 0$, thus
\bea
\lefteqn{ \partial_\tau T^{\tau \tau}
+ \partial_x (v_x T^{\tau \tau})
+ \partial_y (v_y T^{\tau \tau})   } \nonumber\\
&= & - \partial_x \left(v_x P - v_x \pi^{\tau \tau}
+ \pi^{\tau x} \right)
-\partial_y \left(v_y P - v_y \pi^{\tau \tau}
+ \pi^{\tau y} \right) \nonumber \\
&& - \frac{1}{\tau} \left(T^{\tau \tau}
+ P + \tau^2 \pi^{\eta \eta} \right) \, .
\eea
The momentum conservation equations follow from
$\frac{1}{\sqrt{g}}\partial_{\mu} \left(\sqrt{g} \,
T^{\mu i} \right) + \Gamma^{i}_{\mu \beta} T^{\mu \beta} = 0$:
\bea
\lefteqn{\partial_\tau T^{\tau x} + \partial_x (v_x T^{\tau x})
+ \partial_y (v_y T^{\tau x}) =  -\frac{1}{\tau} T^{\tau x}}
\nonumber \\
&-& \partial_x \left(P - v_x \pi^{ \tau x} + \pi^{xx} \right)
- \partial_y \left(- v_y \pi^{ \tau x} + \pi^{xy} \right)
\, , \qquad \\
\lefteqn{\partial_\tau T^{\tau y} + \partial_x (v_x T^{\tau y})
+ \partial_y (v_y T^{\tau y}) = - \frac{1}{\tau} T^{\tau y} }
\nonumber \\
&-& \partial_x \left(- v_x \pi^{\tau y} + \pi^{xy} \right)
- \partial_y \left(P - v_y \pi^{\tau y} + \pi^{yy} \right) \, .
\eea
The use of boost-invariant coordinates affects
the expansion rate,
$\theta_\perp = \gamma_\perp/\tau + \partial_\tau \gamma_\perp
+ \partial_x (\gamma_\perp v_x) + \partial_y (\gamma_\perp v_y)$,
and the $\sigma^{zz}$ component of the shear tensor, which is
replaced by $\sigma^{\eta \eta}
\equiv \tau^{-2} \left(\theta_\perp/3
- \gamma_\perp/\tau\right)$.
The convective time derivative $D$ from Sec.\ \ref{2d}
becomes $D \equiv \gamma_\perp \partial_\tau
+ \gamma_\perp v_x \partial_x + \gamma_\perp v_y \partial_y$.
The relaxation equations are the same as in
Cartesian coordinates except for the replacement
$\pi^{zz} \rightarrow \pi^{\eta \eta}$
which due to a nonvanishing Christoffel symbol
includes a new term, $2 \pi^{\eta \eta} \gamma_\perp/\tau$.
Thus,
\bea
\lefteqn{\gamma_\perp \partial_t \pi^{\eta \eta}
+ \gamma_\perp v_x \partial_x \pi^{\eta \eta}
+ \gamma_\perp v_y \partial_y \pi^{\eta \eta}
= - 2 \pi^{\eta \eta} \frac{\gamma_\perp}{\tau} } \nonumber
\qquad \\
&+& \frac{1}{\tau_\pi}\left(\pi^{\eta \eta}_{NS}
- \pi^{\eta \eta} \right) - I^{\eta \eta}_1  -
I^{\eta \eta}_2 - I^{\eta \eta}_3 \, . \qquad
\eea
Note that in Ref.\ \cite{Heinz:2005bw} this extra term
was not present in Eq.\ (5.21a), but correctly added in
Ref.\ \cite{Song:2007fn}.
The other relaxation equations, together with
$I_0, I^{\mu \nu}_1, I^{\mu \nu}_2$, and $I^{\mu \nu}_3$
and the vorticity tensor components remain formally unchanged.
(This is so, since all nonvanishing Christoffel
symbols are multiplied with $u^{\eta}=0$).

\section{(3+1)--dimensional expansion in Cartesian coordinates}
\label{3d}

This case is very similar to the two-dimensional case
discussed in Sec.\ \ref{2d}.
The only difference is that now the velocity, spatial derivative,
and all four-vector and tensor components in the
$z$ direction are non-zero.
The velocity is $u^{\mu} = \gamma(1,v_x,v_y,v_z)$,
where $\gamma = (1 - v^2_x - v^2_y -v^2_z)^{-1/2}$.
Therefore, the new nonvanishing components of the charge
four-current and energy momentum tensor,
in addition to Eqs.\ (\ref{N02d}) -- (\ref{Ny2d})
and Eqs.\ (\ref{T002d}) -- (\ref{Txy2d}) which formally
remain the same with $\gamma_{\perp} \rightarrow \gamma$, are
\bea
N^{z} &\equiv& N^{0} v_z \, , \\
T^{0z} &\equiv& (e + P )\gamma^2 v_z + \pi^{0z} \, , \\
T^{zz} &\equiv& (e + P)\gamma^2 v^2_z + P  + \pi^{zz} \, ,\\
T^{xz} &\equiv& (e + P)\gamma^2 v_x v_z + \pi^{xz} \, , \\
T^{yz} &\equiv& (e + P)\gamma^2 v_y v_z + \pi^{yz} \, .
\eea
The LRF quantities are calculated similarly
to the (2+1)--dimensional case, thus
\bea
n &=& N^{0} \sqrt{1 - v^2_x - v^2_y - v^2_z} \, ,\\
e &=&  (T^{00} - \pi^{00}) - v_x (T^{0x} - \pi^{0x}) \\ \nonumber
&-& v_y (T^{0y} - \pi^{0y}) - v_z (T^{0z} - \pi^{0z})\, .
\eea
While the velocity components in $x$ and $y$ directions
remain the same as in Sec.\ \ref{2d},
the velocity component in $z$ direction is
\bea \label{vz3d}
v_z = \frac{T^{0z} - \pi^{0z}}{T^{00} - \pi^{00} + P} \, .
\eea
The charge conservation equation is
\bea
\partial_t N^{0} + \partial_x (v_x N^{0})
+ \partial_y (v_y N^{0}) + \partial_z (v_z N^{0}) = 0\, .
\eea
The energy-momentum equations are
\bea
\lefteqn{\partial_t T^{00} + \partial_x (v_x T^{00})
+ \partial_y (v_y T^{00}) + \partial_z (v_z T^{00})}\nonumber \\
& = & - \partial_x \left(v_x P - v_x \pi^{00}
+ \pi^{0x} \right) - \partial_y \left(v_y P - v_y \pi^{00}
+ \pi^{0y} \right) \nonumber \\
& & - \partial_z \left(v_z P - v_z \pi^{00} + \pi^{0z} \right)
\, ,\\
\lefteqn{\partial_t T^{0x} + \partial_x (v_x T^{0x})
+  \partial_y (v_y T^{0x}) + \partial_z (v_z T^{0x})}\nonumber \\
&=& - \partial_x \left(P - v_x \pi^{0x} + \pi^{xx} \right)
- \partial_y \left(- v_y \pi^{0x} + \pi^{xy} \right) \nonumber \\
& & - \partial_z \left(- v_z \pi^{0x} + \pi^{xz} \right)
\, ,\\
\lefteqn{\partial_t T^{0y} + \partial_x (v_x T^{0y})
+ \partial_y (v_y T^{0y}) + \partial_z (v_z T^{0y}) }\nonumber \\
&=& - \partial_x \left(- v_x \pi^{0y} + \pi^{xy} \right)
- \partial_y \left(P - v_y \pi^{0y} + \pi^{yy} \right)
\nonumber \\
& & - \partial_z \left(- v_z \pi^{0y} + \pi^{yz} \right)
\, ,\\
\lefteqn{\partial_t T^{0z} + \partial_x (v_x T^{0z})
+ \partial_y (v_y T^{0z}) + \partial_z (v_z T^{0z})}\nonumber \\
&=& - \partial_x \left(- v_x \pi^{0z} + \pi^{xz} \right)
- \partial_y \left(- v_y \pi^{0z} + \pi^{yz} \right) \nonumber \\
& & -\partial_z \left(P - v_z \pi^{0z} + \pi^{zz} \right) \, .
\eea
The relaxation equations are formally similar to
Eqs.\ (\ref{relax_bulk2d}), (\ref{relax_shear2d}),
only the $z$-directed derivatives
$\gamma v_z \partial_z \Pi$ and
$\gamma v_z \partial_z \pi^{\mu\nu}$
have to be added.
Therefore, the new components of the shear tensor are
\bea
\sigma^{0z}
&=& \frac{1}{2}\left[\partial_t (\gamma v_z)
- \partial_z \gamma \right] \nonumber \\
&-& \frac{1}{2}\left[\gamma D (\gamma v_z)
+ \gamma v_z D\gamma \right] + \gamma^2 v_z \frac{\theta}{3}
\, ,\\
\sigma^{zz}
&=& - \partial_z (\gamma v_z) - \gamma v_z D (\gamma v_z)
+ (1 + \gamma^2 v^2_z) \frac{\theta}{3} \, ,\\
\sigma^{xz}
&=& -\frac{1}{2} \left[ \partial_x (\gamma v_z)
+ \partial_z (\gamma v_x) \right]  \nonumber \\
&-& \frac{1}{2} \left[ \gamma v_x D (\gamma v_z)
+ \gamma v_z D (\gamma v_x) \right]
+ \gamma^2 v_x v_z \frac{\theta}{3} \, , \\
\sigma^{yz}
&=& -\frac{1}{2} \left[ \partial_y (\gamma v_z)
+ \partial_z (\gamma v_y) \right] \nonumber \\
&-& \frac{1}{2} \left[ \gamma v_y D (\gamma v_z)
+ \gamma v_y D (\gamma v_z) \right]
+ \gamma^2 v_y v_z \frac{\theta}{3} \, , \qquad
\eea
where the expansion scalar is
$\theta = \partial_t \gamma + \partial_x (\gamma v_x)
+ \partial_y (\gamma v_y) + \partial_z (\gamma v_z)$,
and the convective time derivative is
$D \equiv u^{\mu} \partial_{\mu} = \gamma \partial_t
+ \gamma v_x \partial_x + \gamma v_y \partial_y
+ \gamma v_z \partial_z$.
The form of the other components does not change in comparison
with Eqs.\ (\ref{s002d}) -- (\ref{sxy2d}).

The term
$I^{\mu \nu}_1 = (\pi^{\lambda\mu}u^\nu+\pi^{\lambda\nu}u^\mu)
D u_\lambda$ leads to
\bea \nonumber
I^{00}_1  &=&  2 \gamma \left[\pi^{00} D \gamma
- \pi^{0x} D (\gamma v_x) - \pi^{0y} D (\gamma v_y) \right. \\
&-& \left. \pi^{0z} D (\gamma v_z) \right] \, , \\ \nonumber
I^{0x}_1  &=& \gamma \left[ (\pi^{00} v_x
+ \pi^{0x})D \gamma - (\pi^{0x} v_x  + \pi^{xx})D (\gamma v_x)
\right. \\ \nonumber
&-& \left. (\pi^{0y} v_x  + \pi^{xy})D (\gamma v_y) \right. \\
&-& \left. (\pi^{0z} v_x  + \pi^{xz})D (\gamma v_z) \right]\, ,
\\ \nonumber
I^{0y}_1 \! \! &=& \! \! \gamma \left[
(\pi^{00} v_y  + \pi^{0y})D \gamma -
(\pi^{0x} v_y  + \pi^{xy})D (\gamma v_x) \right. \\ \nonumber
&-& \left. (\pi^{0y} v_y  + \pi^{yy})D (\gamma v_y) \right. \\
&-& \left. (\pi^{0z} v_y  + \pi^{yz})D (\gamma v_z) \right]\, ,
\\ \nonumber
I^{0z}_1 \! \! &=& \! \!  \gamma \left[
(\pi^{00} v_z  + \pi^{0z})D \gamma -
(\pi^{0x} v_z  + \pi^{xz})D (\gamma v_x) \right. \\ \nonumber
&-& \left. (\pi^{0y} v_z  + \pi^{yz})D (\gamma v_y) \right. \\
&-& \left. (\pi^{0z} v_z  + \pi^{zz})D (\gamma v_z) \right]
\, , \\ \nonumber
I^{xx}_1 \! \! &=& \! \!  2 \gamma v_x \left[
\pi^{0x} D \gamma  - \pi^{xx} D (\gamma v_x) -
\pi^{xy} D (\gamma v_y) \right. \\
&-& \left. \pi^{xz} D (\gamma v_z) \right]\, , \\ \nonumber
I^{yy}_1 \! \! &=& \! \!  2 \gamma v_y \left[ \pi^{0y}
D \gamma  - \pi^{xy} D (\gamma v_x) - \pi^{yy} D (\gamma v_y)
\right. \\
&-& \left. \pi^{yz} D (\gamma v_z) \right]\, , \\ \nonumber
I^{zz}_1 \! \! &=& \! \!  2 \gamma v_z \left[
\pi^{0z} D \gamma  - \pi^{xz} D (\gamma v_x)
- \pi^{yz} D (\gamma v_y) \right. \\
&-& \left. \pi^{zz} D (\gamma v_z) \right]\, , \\ \nonumber
I^{xy}_1 \! \! &=& \! \!\gamma \left[
(\pi^{0x} v_y  + \pi^{0y}v_x)D \gamma
- (\pi^{xx} v_y  + \pi^{xy}v_x)D (\gamma v_x)
\right. \\ \nonumber
&-& \left. (\pi^{xy} v_y  + \pi^{yy}v_x)D (\gamma v_y) \right. \\
&-& \left. (\pi^{xz} v_y  + \pi^{yz}v_x)D (\gamma v_z) \right]
 \, , \\ \nonumber
I^{xz}_1 \! \! &=& \! \! \gamma \left[
(\pi^{0x} v_z  + \pi^{0z}v_x)D \gamma
- (\pi^{xx} v_z  + \pi^{xz}v_x)D (\gamma v_x)
\right. \\ \nonumber
&-& \left. (\pi^{xy} v_z  + \pi^{yz}v_x)D (\gamma v_y) \right. \\
&-& \left. (\pi^{xz} v_z  + \pi^{zz}v_x)D (\gamma v_z) \right]
\, , \\ \nonumber
I^{yz}_1 \! \! &=& \! \! \gamma \left[
(\pi^{0y} v_z  + \pi^{0z}v_y)D \gamma
- (\pi^{xy} v_z  + \pi^{xz}v_y)D (\gamma v_x)
\right. \\ \nonumber
&-& \left. (\pi^{yy} v_z  + \pi^{yz}v_y)D (\gamma v_y) \right. \\
&-& \left. (\pi^{yz} v_z  + \pi^{zz}v_y)D (\gamma v_z) \right]
\, .
\eea
The terms $I_0$ and $I^{\mu \nu}_2$ are given by
Eqs.\ (\ref{I_0}) and (\ref{I_2}).
The new components which need to be computed
compared to the previous case are
$I^{0z}_2,I^{xz}_2,I^{yz}_2$, and $I^{zz}_2$.
The non-vanishing components of the last term are
\bea
I^{00}_3 &=& 2\left(\pi^{0x} \omega^0_{\hspace*{0.1cm} x}
+ \pi^{0y} \omega^0_{\hspace*{0.1cm} y}
+ \pi^{0z} \omega^0_{\hspace*{0.1cm} z}\right) \, , \\ \nonumber
I^{0x}_3 &=& \pi^{00} \omega^x_{\hspace*{0.1cm} 0}
+ \pi^{0y} \omega^x_{\hspace*{0.1cm} y}
+ \pi^{0z} \omega^x_{\hspace*{0.1cm} z}  \\
&+& \pi^{xx} \omega^0_{\hspace*{0.1cm} x}
+ \pi^{xy} \omega^0_{\hspace*{0.1cm} y}
+ \pi^{xz} \omega^0_{\hspace*{0.1cm} z} \, ,\\ \nonumber
I^{0y}_3 &=& \pi^{00} \omega^y_{\hspace*{0.1cm} 0}
+ \pi^{0x} \omega^y_{\hspace*{0.1cm} x}
+ \pi^{0z} \omega^y_{\hspace*{0.1cm} z} \\
&+& \pi^{xy} \omega^0_{\hspace*{0.1cm} x}
+ \pi^{yy} \omega^0_{\hspace*{0.1cm} y}
+ \pi^{yz} \omega^0_{\hspace*{0.1cm} z} \, ,\\ \nonumber
I^{0z}_3 &=& \pi^{00} \omega^z_{\hspace*{0.1cm} 0}
+ \pi^{0x} \omega^z_{\hspace*{0.1cm} x}
+ \pi^{0y} \omega^z_{\hspace*{0.1cm} y} \\
&+& \pi^{xz} \omega^0_{\hspace*{0.1cm} x}
+ \pi^{yz} \omega^0_{\hspace*{0.1cm} y}
+ \pi^{zz} \omega^0_{\hspace*{0.1cm} z} \, , \\
I^{xx}_3 &=& 2 \left( \pi^{0x} \omega^x_{\hspace*{0.1cm} 0}
+ \pi^{xy} \omega^x_{\hspace*{0.1cm} y}
+ \pi^{xz} \omega^x_{\hspace*{0.1cm} z} \right) \, , \\
I^{yy}_3 &=& 2 \left( \pi^{0y} \omega^y_{\hspace*{0.1cm} 0}
+ \pi^{xy} \omega^y_{\hspace*{0.1cm} x}
+ \pi^{yz} \omega^y_{\hspace*{0.1cm} z} \right) \, , \\
I^{zz}_3 &=& 2 \left( \pi^{0z} \omega^z_{\hspace*{0.1cm} 0}
+ \pi^{xz} \omega^z_{\hspace*{0.1cm} x}
+ \pi^{yz} \omega^z_{\hspace*{0.1cm} y} \right) \, , \\ \nonumber
I^{xy}_3 &=& \pi^{0x} \omega^y_{\hspace*{0.1cm} 0}
+ \pi^{xx} \omega^y_{\hspace*{0.1cm} x}
+ \pi^{xz} \omega^y_{\hspace*{0.1cm} z} \\
&+& \pi^{0y} \omega^x_{\hspace*{0.1cm} 0}
+ \pi^{yy} \omega^x_{\hspace*{0.1cm} y}
+ \pi^{yz} \omega^x_{\hspace*{0.1cm} z} \, , 
\eea
\bea \nonumber
I^{xz}_3 &=& \pi^{0x} \omega^z_{\hspace*{0.1cm} 0}
+ \pi^{xx} \omega^z_{\hspace*{0.1cm} x}
+ \pi^{xy} \omega^z_{\hspace*{0.1cm} y} \\
&+& \pi^{0z} \omega^x_{\hspace*{0.1cm} 0}
+ \pi^{yz} \omega^x_{\hspace*{0.1cm} y}
+ \pi^{zz} \omega^x_{\hspace*{0.1cm} z} \, , \\ \nonumber
I^{yz}_3 &=& \pi^{0y} \omega^z_{\hspace*{0.1cm} 0}
+ \pi^{xy} \omega^z_{\hspace*{0.1cm} x}
+ \pi^{yy} \omega^z_{\hspace*{0.1cm} y} \\
&+& \pi^{0z} \omega^y_{\hspace*{0.1cm} 0}
+ \pi^{xz} \omega^y_{\hspace*{0.1cm} x}
+ \pi^{zz} \omega^y_{\hspace*{0.1cm} z} \, ,
\eea
where the new vorticity tensor components are
\bea
\omega^0_{\hspace*{0.1cm} z} &=&
\frac{1}{2}\left[\partial_z \gamma + \partial_t (\gamma v_z)
+ \gamma v_z D \gamma - \gamma D (\gamma v_z)\right]
\, , \qquad \\ \nonumber
\omega^x_{\hspace*{0.1cm} z} &=&
\frac{1}{2} \left[ \partial_z (\gamma v_x)
- \partial_x (\gamma v_z) \right] \\
&+& \frac{1}{2} \left[ \gamma v_z D (\gamma v_x)
- \gamma v_x D (\gamma v_z) \right] \, , \\ \nonumber
\omega^y_{\hspace*{0.1cm} z} &=&
\frac{1}{2} \left[ \partial_z (\gamma v_y)
- \partial_y (\gamma v_z) \right] \\
&+& \frac{1}{2} \left[ \gamma v_z D (\gamma v_y)
- \gamma v_y D (\gamma v_z) \right] \, ,
\eea
such that $\omega^0_{\hspace*{0.1cm} z} =
\omega^z_{\hspace*{0.1cm} 0} = -\omega^{0z} = \omega_{0z}$,
$\omega^x_{\hspace*{0.1cm} z} = -\omega^z_{\hspace*{0.1cm} x}
= -\omega^{xz} = -\omega_{xz}$ and
$\omega^y_{\hspace*{0.1cm} z} = -\omega^z_{\hspace*{0.1cm} y}
= -\omega^{yz} = -\omega_{yz} $.
The other components are given in Eqs.\
(\ref{w0x2d}) -- (\ref{wxy2d}) where one has to
replace $\gamma_\perp$ with $\gamma$.

\section{(3+1)--dimensional expansion in  $(\tau, x, y, \eta)$ coordinates}
 \label{3d_lc}

The metric of the space-time is the same as in
App.\ \ref{2d_lc}, only the definition of the flow velocity
changes. In this case the contravariant flow velocity is
$u^{\mu} = \gamma(1,v_x,v_y,v_\eta)$, and the
covariant flow velocity is
$u_{\mu} = g_{\mu \nu} u^{\nu} =
\gamma(1,-v_x,-v_y,-\tau^2 v_\eta)$,
where $\gamma = (1 - v^2_x - v^2_y - \tau^2 v^2_\eta)^{-1/2}$.
The gradients are, $\partial_\mu
= (\partial_\tau, \partial_x, \partial_y, \partial_\eta)$ and
$\partial^\mu \equiv g^{\mu \nu} \partial_{\nu}
= (\partial_\tau, -\partial_x, -\partial_y, -\tau^{-2}
\partial_\eta)$.

Similarly as before the equations can be easily obtained from
the ones found in Cartesian coordinates in App.\ \ref{3d}.
All laboratory frame quantities are formally the same, except
for $T^{zz} \rightarrow T^{\eta \eta}
\equiv (e + P)\gamma^2 v^2_\eta  + P/\tau^2 + \pi^{\eta \eta}$.
The LRF quantities are
\bea
n &=& N^{0} \sqrt{1 - v^2_x - v^2_y - \tau^2 v^2_\eta}
 \, ,\\ \nonumber
e &=&  (T^{00} - \pi^{00}) - v_x (T^{0x} - \pi^{0x}) \\
&-& v_y (T^{0y} - \pi^{0y}) - \tau^2 v_\eta (T^{0\eta}
- \pi^{0\eta})\, .
\eea
The velocity components $v_x$ and $v_y$ are given by
Eqs.\ (\ref{vx2d}), (\ref{vy2d}), the velocity component in
$\eta$-direction is given similarly as in Eq.\ (\ref{vz3d}).
The charge conservation equation is given by
\bea
\partial_\tau N^{\tau} &+& \partial_x (v_x N^{\tau})
+ \partial_y (v_y N^{\tau}) + \partial_\eta (v_\eta N^{\tau})
\nonumber \\
&=&  -\frac{1}{\tau} N^{\tau}\, . \qquad
\eea
The equation for energy-momentum conservation leads to
\bea
\lefteqn{\partial_\tau T^{\tau \tau}
+ \partial_x (v_x T^{\tau \tau})
+ \partial_y (v_y T^{\tau \tau})
+ \partial_\eta (v_\eta T^{\tau \tau})
}\nonumber \\ \nonumber
&=& - \partial_x \left(v_x P - v_x \pi^{\tau \tau}
+ \pi^{\tau x} \right)
-\partial_y \left(v_y P - v_y \pi^{\tau \tau}
+ \pi^{\tau y} \right) \\
& &- \partial_\eta \left(v_\eta P - v_\eta \pi^{\tau \tau}
+ \pi^{\tau \eta} \right) - \frac{1}{\tau} \left(T^{\tau \tau}
+ \tau^2 T^{\eta \eta} \right) \, , \\
\lefteqn{\partial_\tau T^{\tau x} + \partial_x (v_x T^{\tau x})
+ \partial_y (v_y T^{\tau x})
+ \partial_\eta (v_\eta T^{\tau x})
}\nonumber \\ \nonumber
&=& - \partial_x \left(P - v_x \pi^{ \tau x} + \pi^{xx} \right)
- \partial_y \left(- v_y \pi^{ \tau x} + \pi^{xy} \right) \\
& & -\partial_y \left(- v_\eta \pi^{ \tau x}
+ \pi^{x\eta} \right)
- \frac{1}{\tau} T^{\tau x} \, , \\
\lefteqn{\partial_\tau T^{\tau y} + \partial_x (v_x T^{\tau y})
+ \partial_y (v_y T^{\tau y})
+ \partial_\eta (v_\eta T^{\tau y})
}\nonumber \\ \nonumber
&=& - \partial_x \left(- v_x \pi^{\tau y} + \pi^{xy} \right)
- \partial_y \left(P - v_y \pi^{\tau y} + \pi^{yy} \right) \\
& &- \partial_\eta \left(- v_\eta \pi^{\tau y}
+ \pi^{y\eta} \right)
- \frac{1}{\tau} T^{\tau y}  \, , \\
\lefteqn{\partial_\tau T^{\tau \eta}
+ \partial_x (v_x T^{\tau \eta})
+ \partial_y (v_y T^{\tau \eta})
+ \partial_\eta (v_\eta T^{\tau \eta})
} \nonumber \\ \nonumber
&= &- \partial_x \left(- v_x \pi^{\tau \eta}
+ \pi^{x\eta} \right)
- \partial_y \left(- v_y \pi^{\tau \eta} + \pi^{y\eta} \right) \\
& & - \partial_\eta \left(\frac{P}{\tau^2}
- v_\eta \pi^{\tau \eta}
+ \pi^{\eta \eta} \right) - \frac{3}{\tau} T^{\tau \eta}  \, .
\eea
The relaxation equations for the bulk viscous pressure and
the shear stress tensor components
$\pi^{xx},\pi^{yy},\pi^{xy}$ are formally the same
as in Cartesian coordinates, however, for the other
components we obtain
\bea
D \pi^{\tau \tau} &= &- 2 \tau \gamma v_\eta \pi^{\tau \eta}
+  I^{\tau \tau} \, , \\
 D\pi^{\tau x} &=&  - \tau \gamma v_\eta \pi^{x \eta}
+ I^{\tau x} \, , \\
 D \pi^{\tau y}& =& - \tau \gamma v_\eta \pi^{y \eta}
+ I^{\tau y}\, , \\
D \pi^{\tau \eta} &=& - \tau \gamma v_\eta \pi^{\eta \eta}
- \frac{\gamma}{\tau} \pi^{\tau \eta}
- \frac{\gamma v_\eta}{\tau} \pi^{\tau \tau} + I^{\tau \eta}
\, , \qquad \\
D \pi^{x \eta} & =& - \frac{\gamma}{\tau} \pi^{x \eta}
- \frac{\gamma v_\eta}{\tau} \pi^{\tau x}
+ I^{x \eta} \, , \\
D \pi^{y \eta} &=& - \frac{\gamma}{\tau} \pi^{y \eta}
- \frac{\gamma v_\eta}{\tau} \pi^{\tau y}
+ I^{y \eta} \, ,\\
D \pi^{\eta \eta}& =& - 2\frac{\gamma}{\tau} \pi^{\eta \eta}
- 2\frac{\gamma v_\eta}{\tau} \pi^{\tau \eta}
+ I^{\eta \eta}\, .
\eea
where $I^{\mu \nu}$ denotes the right-hand
side of Eq.\ (\ref{relax_shear}),
but in this case
$D = \gamma\partial_\tau + \gamma v_x \partial_x
+ \gamma v_y\partial_y  + \gamma v_\eta \partial_\eta$
denotes the convective time derivative of scalars.

The shear tensor components $\sigma^{xx}, \sigma^{yy}$,
and $\sigma^{xy}$ remain formally unchanged from
Eqs.\ (\ref{sxx2d}), (\ref{syy2d}), (\ref{sxy2d}), while
the ones which are different are calculated
from Eq.\ (\ref{shear}),
\bea
\sigma^{\tau \tau}
&=& - \tau \gamma^3 v^2_\eta + \partial_\tau \gamma
- \gamma D\gamma
+ \left(\gamma^2 - 1\right)\frac{\theta}{3} \, ,\qquad
\eea
\bea \nonumber
\sigma^{\tau x}
&=& - \frac{\tau \gamma^3 v^2_\eta v_x}{2}
+ \frac{1}{2} \left[ \partial_\tau (\gamma v_x)
- \partial_x \gamma \right] \\
&-& \frac{1}{2} \left[\gamma D (\gamma v_x)
+ \gamma v_x D\gamma \right]
+ \gamma^2 v_x \frac{\theta}{3} \, ,\\ \nonumber
\sigma^{\tau y}
&=& - \frac{\tau \gamma^3 v^2_\eta v_y}{2}
+ \frac{1}{2}\left[\partial_\tau (\gamma v_y)
- \partial_y \gamma \right] \\
&-& \frac{1}{2}\left[\gamma D (\gamma v_y)
+ \gamma v_y D\gamma \right]
+ \gamma^2 v_y \frac{\theta}{3} \, ,\\ \nonumber
\sigma^{\tau \eta}
&=& - \frac{\gamma^3 v_\eta}{2\tau} \left(2
+ \tau^2 v^2_\eta \right)
+ \frac{1}{2}\left[\partial_\tau (\gamma v_\eta)
- \frac{1}{\tau^2} \partial_\eta \gamma \right] \qquad \\
&-& \frac{1}{2}\left[\gamma D (\gamma v_\eta)
+ \gamma v_\eta D\gamma \right]
+ \gamma^2 v_\eta \frac{\theta}{3} \, ,\\ \nonumber
\sigma^{\eta \eta}
&=& -\frac{\gamma}{\tau^3} \left(1
+ 2 \gamma^2 v^2_\eta \tau^2 \right)
- \frac{1}{\tau^2}\partial_\eta (\gamma v_\eta)  \\
&-& \gamma v_\eta D (\gamma v_\eta) + (\frac{1}{\tau^2}
+ \gamma^2 v^2_\eta) \frac{\theta}{3} \, ,\\ \nonumber
\sigma^{x\eta}
&=& -\frac{\gamma^3 v_x v_\eta}{\tau}
- \frac{1}{2} \left[ \partial_x (\gamma v_\eta)
+ \frac{1}{\tau^2}\partial_\eta (\gamma v_x) \right] \qquad \\
&-& \frac{1}{2} \left[ \gamma v_x D (\gamma v_\eta)
+ \gamma v_\eta D (\gamma v_x) \right]
+ \gamma^2 v_x v_\eta \frac{\theta}{3} \, , \\ \nonumber
\sigma^{y\eta}
&=& -\frac{\gamma^3 v_y v_\eta}{\tau}
- \frac{1}{2} \left[ \partial_y (\gamma v_\eta)
+ \frac{1}{\tau^2}\partial_\eta (\gamma v_y) \right] \\
&-& \frac{1}{2} \left[ \gamma v_y D (\gamma v_\eta)
+ \gamma v_\eta D (\gamma v_y) \right]
+ \gamma^2 v_y v_\eta \frac{\theta}{3} \, ,
\eea
where the expansion scalar is
$\theta = \gamma/\tau + \partial_t \gamma
+ \partial_x (\gamma v_x) + \partial_y (\gamma v_y)
+ \partial_\eta (\gamma v_\eta)$.
The $I_0$, $I^{\mu \nu}_1$, $I^{\mu \nu}_2$, and
$I^{\mu \nu}_3$ components remain formally the same.
The new vorticity tensor components are
\bea  \nonumber
\omega^\tau_{\hspace*{0.1cm} \eta}
&=& \frac{1}{2}\left[\partial_\eta \gamma
+ \partial_\tau (\tau^2 \gamma v_\eta) \right] \\
&+& \frac{1}{2}\left[\tau^2\gamma v_\eta D \gamma
- \gamma D (\tau^2 \gamma v_\eta)\right] \, , \\  \nonumber
\omega^x_{\hspace*{0.1cm} \eta} &=&
\frac{1}{2} \left[ \partial_\eta (\gamma v_x)
- \partial_x (\tau^2 \gamma v_\eta) \right] \\
&+& \frac{1}{2} \left[\tau^2 \gamma v_\eta D (\gamma v_x)
- \gamma v_x D (\tau^2 \gamma v_\eta) \right] \, , \\ \nonumber
\omega^y_{\hspace*{0.1cm} \eta} &=&
\frac{1}{2} \left[ \partial_\eta (\gamma v_y)
- \partial_y (\tau^2 \gamma v_\eta) \right] \\
&+& \frac{1}{2} \left[ \tau^2 \gamma v_\eta D (\gamma v_y)
- \gamma v_y D (\tau^2 \gamma v_\eta) \right] \, ,
\eea
where $\omega^{\tau}_{\hspace*{0.1cm} \eta}
= \omega^{\eta}_{\hspace*{0.1cm} \tau}$,
$\omega^{x}_{\hspace*{0.1cm} \eta}
= -\omega^{\eta}_{\hspace*{0.1cm} x}$ and
$\omega^{y}_{\hspace*{0.1cm} \eta}
= -\omega^{\eta}_{\hspace*{0.1cm} y}$.




\end{document}